\documentclass[twocolumn]{aastex631}

\newcommand{\hi}{H {\sc i} }
\newcommand{\civ}{C {\sc iv}}
\newcommand{\heii}{He {\sc ii}}

\newcommand{\mgii}{Mg {\sc ii}}
\newcommand{\feii}{Fe {\sc ii}}

\shorttitle{AASTeX v6.3.1 Sample article}
\shortauthors{Zhang et al.}
\usepackage{color}
\usepackage{amsmath}
\usepackage{amssymb}
\usepackage{float}
\usepackage{changepage}
\usepackage{multirow}
\graphicspath{{./}{figures/}}

\begin{document}

\title{Revealing the Gas Recycling in the Circumgalactic Medium (CGM) Utilizing a Luminous Ly$\alpha$ nebula around a Type-II Quasar at $z=2.6$ with the Keck Cosmic Web Imager (KCWI)}

\author{Shiwu Zhang}
\affiliation{Department of Astronomy, Tsinghua University, Beijing 100084, China}
\author{Zheng Cai}
\affiliation{Department of Astronomy, Tsinghua University, Beijing 100084, China}
\author{Dandan Xu}
\affiliation{Department of Astronomy, Tsinghua University, Beijing 100084, China}
\author{Andrea Afruni}
\affiliation{Departamento de Astronomía, Universidad de Chile, Casilla 36-D, Santiago, Chile}
\author{Yunjing Wu}
\affiliation{Department of Astronomy, Tsinghua University, Beijing 100084, China}
\affiliation{Steward Observatory, University of Arizona, 933 N Cherry Ave, Tucson, AZ 85721, USA}
\author{Wuji Wang}
\affiliation{Astronomisches Rechen-Institut, Zentrum f{\"u}r Astronomie der Universit{\"a}t Heidelberg, M{\"o}nchhofstr. 12-14, 69120 Heidelberg, Germany}
\author{Fabrizio Arrigoni Battaia}
\affiliation{Max-Planck-Institut für Astrophysik, Karl-Schwarzschild-Str 1, D-85748 Garching bei München, Germany}
\author{Mingyu Li}
\affiliation{Department of Astronomy, Tsinghua University, Beijing 100084, China}
\author{Sen Wang}
\affiliation{Department of Astronomy, Tsinghua University, Beijing 100084, China}
\author{Xianzhi Bi}
\affiliation{Beijing Royal School, Beijing 102209, China}







\begin{abstract}
How galaxies acquire material from the circumgalactic medium (CGM) is a key question in galaxy evolution.
Recent observations and simulations show that gas recycling could be an important avenue for star formation. 
This paper presents Keck Cosmic Web Imager (KCWI) integral field unit spectroscopic observations on a type-II quasar, Q1517+0055 at $z=2.65$, a pilot study of our Ly$\alpha$ nebulae sample at $z\approx2$.
We revealed diffuse emission of the Ly$\alpha$ $1216$, \heii \ $1640$, and \civ \ $1549$ on the projected physical scale of 122 kpc, 45 kpc, and 79 kpc, respectively. 
The total Ly$\alpha$ luminosity is $L_{\rm Ly\alpha}= 3.04\pm 0.02 \times 10^{44} \ {\rm erg \ s^{-1}}$.
The line ratio diagnostics 
shows that ${\rm HeII/Ly\alpha}\approx 0.08$ and ${\rm CIV/Ly\alpha}\approx 0.28$, consistent with the photoionization including recombination and photon punmping.
We also identify the associated \hi and \civ \ absorption from the spectra. 
By fitting the spectra, we derive both the column density and the velocity. 
We find that the velocity profile from both the absorption and the \heii \ emission exhibit increasing trends. 
Moreover, both the line ratio diagnostic from the emission and the column density ratio from the absorption confirm that the cool gas metallicity is $\geq Z_{\odot}$.
From detailed modeling and estimation, gas recycling might be a more plausible interpretation compared with the scenario of a powerful outflow.
\end{abstract}


\section{Introduction} \label{intro}

The gas diffusing in the dark matter halo is the circumgalactic medium (CGM) which plays a key role in galaxy formation and evolution \citep{Tumplison2017}.
As the link between the intergalactic medium (IGM) and galaxies, CGM has been confirmed to contain multi-phase gas including the hot gas ($T\geq 10^{5}$ K) observed by the X-ray emission \citep{Anderson2016, Anderson2013}, the cool gas ($T\sim 10^{4}$ K) observed by the UV-optical line emissions \citep{Cai2017, Cai2019, FAB2019,Borisova2016}, and the cold gas ($T\sim 10-100$ K) observed by radio/submillimeter emissions \citep{Emonts2018, Emonts2019}.
Hydrodynamical simulations \citep{kere2005, angle2017, suresh2015,Wangsen2021,Lu2022} have shown that the mass, energy, and metals exchange between the galaxy and the CGM/IGM is complex.
Despite the great theoretical and observational efforts, 
the origin, the cooling mechanisms, and the detailed dynamical processes of the CGM are 
still under investigation.

In the modern paradigm of how galaxies obtain their gas, the well-accepted picture is that the galaxy could not only accrete the hot gas isotropically \citep{nelson2013,stern2020} but also obtain the cool gas through filaments in the form of inspiraling stream \citep{kere2005, Dekel2009}. 
Nevertheless, beyond this sketch, studies indicate that gas recycling could also be an important avenue for the galaxy to sustain star formation \citep{angle2017,oppenheimer2010}.
In the so-called recycling scenario, the metal-enriched gas is firstly ejected to the CGM by the outflow, and then falls back to the galaxy \citep{ford2014,Christensen2016}. 
This phenomenon has been studied in both observations and simulations at low-$z$ \citep{Rubin2012,oppenheimer2010}.
Hydrodynamical simulations show that the gas recycling could exceed the cold-mode accretion at $z=0$ and reshape the stellar mass function (SMF) of the galaxy \citep{oppenheimer2010}. 
\cite{Rubin2012} present the detection of the metal-enriched accretion with the \mgii \ and \feii \ absorption at $z\leq1$. 
For high-$z$, the cosmological simulations show that recycled gas is an important source to regulate star formation. 
FIRE simulations show that the recycled gas can take a fraction more than 50\% of the total accreted material at $z\approx2$ \citep{angle2017}.
Also, \cite{Wangsen2021} using IllustrisTNG simulations demonstrate a closely entangled fate between the CGM gas recycling and episodic star formation. 
Nevertheless, the direct detection of gas recycling at high-$z$ is rare.


Ly$\alpha$ nebulae 
are good laboratories allowing us to directly study the cool gas in the CGM.
By observing the Enormous Ly$\alpha$ nebula (ELAN), MAMMOTH-1, \cite{zhang2022} present the first direct imaging on the recycled inflow in a massive system with halo mass of $10^{13} M_{\odot}$ at $z=2.3$.
They find that the cool gas is enriched to $Z_{\odot}$ by the previous outflow, which is flowing back to the galaxy in the form of inspiraling streams.
Furthermore, the gas inflow rate is estimated to be $\dot{M}_{\rm in}\approx 713 \ M_{\odot}$ yr$^{-1}$, fully covering the star formation rate (SFR) of the central galaxy.
These observations also indicate that the recycled inflow could be important for enriching the star-forming environment in massive systems at high-$z$.

Here, we present a pilot study targeting a type-II quasar, Q1517+0555, at $z=2.65$ with the Keck Cosmic Web Imager (KCWI) instrument on Keck Telescope.
This source is selected from a type-II quasar candidate catalog \citep{alexandroff2013}.
After 2-hr exposure with Keck/KCWI which is a blue-sensitive integral field spectrograph (IFS),
we detect not only extended Ly$\alpha$ emission line, but also extended \civ \ and \heii\ emission lines. 
Besides, we also find the \hi\ and \civ\ absorption associated with the emission.
After analyzing the absorption, we conclude that the absorption could trace the metal-enriched cool gas inflowing into the galaxy.

This paper is organized as follows: we present the information of observations in Sec.~\ref{observations} where the details about the systemic redshift of the source and the process of the data reduction are also given.
In Sec.~\ref{results}, we show the detailed results of Ly$\alpha$, \heii, and \civ \ emissions with the \hi \ and \civ \ absorption.
In Sec.~\ref{discussion}, the possible powering mechanism and the possible explanations of our observations are discussed.
Conclusions are provided in Sec.~\ref{conclusion}.
The $\Lambda$CDM cosmology with $\Omega_{\rm m}=0.3$, $\Omega_{\Lambda}=0.7$, and $h=0.7$ is assumed.

\section{Observations} \label{observations}
\subsection{KCWI observations} \label{sec:KCWI}

We obtained the IFS observations of the Q1517+0055 with the KCWI mounted on the Keck-II telescope \citep{Morrissey2018}.
The medium slicer with the field of view (FoV) of $16''\times20''$ was employed for observations. 
It yields a spatial resolution of $0.7''$ along the slicer direction with the seeing-limited condition of $0.6''$. 
The KCWI/BM grating with the wavelength range of $3500 \ {\rm \AA} - 6200\ {\rm \AA}$ was employed.
This configuration yields a slit-limited spectral resolution of $R\approx4000$, corresponding to a velocity resolution with the full width of half maximum (FWHM) of 75 km s$^{-1}$ (rest-frame 22.7 km s$^{-1}$), completely resolving the kinematics of the cool gas of the Ly$\alpha$ nebula.

To fully reveal the morphology and kinematics, we performed a 2-hours on-source exposure. 
This yields a 2-$\sigma$ surface brightness (SB) of $8.4\times 10^{-19}$ erg s$^{-1}$ cm$^{-2}$ arcsec$^{-2}$ at $\lambda \approx4400 \ {\rm \AA}$, assuming a wavelength bin of $1 \ {\rm \AA}$. 
At $\lambda=5800 \ {\rm \AA}$, our observations yields the 2-$\sigma$ surface brightness of $6.5\times 10^{-19}$ erg s$^{-1}$ cm$^{-2}$ arcsec$^{-2}$ in a bin of $1 \ {\rm \AA}$.



\subsection{Data Reduction} \label{sec:reduction}
The standard KCWI pipeline \footnote{https://github.com/Keck-DataReductionPipelines/KcwiDRP} was employed to reduce the data. 
We first subtract the bias, correct the pixel-to-pixel variation with the flat-field images, and remove the cosmic rays.
Then the geometric transformation and the wavelength calibration with ThAr arc images are conducted.
We use the twilight flats to correct the slice-to-slice variance. 
The spectrophotometric standard star is adopted to calibrate the flux of the individual image of the cube.

The systemic redshift is calculated from the \heii \ line which is  non-resonant. 
By fitting the \heii \ emission, we have $\lambda_{\rm HeII}=5983.2 \pm 0.3$ \AA, corresponding to a systemic redshift of $z_{\rm sys}=2.6474 \pm 0.0002$ with the rest-frame wavelength of \heii \ equal to $1640.4 \ {\rm \AA}$.

The optimal-extraction method for faint, extended sources extracting \citep{Borisova2016,FAB2019,Cai2019} is applied to generate the pseudo narrowband images for the Ly$\alpha$, \heii, and \civ \ emissions.
We extract the sub-cubes by centering on the redshifted line centers and fixing the wavelength bin size of $\Delta \lambda=\pm 20 \ {\rm \AA}$. 
With the systemic redshift of $z_{\rm sys}=2.6474\pm 0.0002$, the wavelength coverage of the sub-cubes are fixed to be $4410 - 4450 \ {\rm \AA}$, $5963 - 6003 \ {\rm \AA}$, and $5631 - 5671 \ {\rm \AA}$. 
Then, we apply the Gaussian kernel with ${\rm FWHM=0.7''}$, same as the spatial resolution along the slicer direction, to smooth the sub-cubes. 
We construct the three-dimensional segmentation masks, containing the value of zero or one to select the connected voxels which have the $S/N$ above the user-defined thresholds.
a threshold of $S/N=2$ is employed.
Such threshold has been tested extensively in previous IFS studies \citep{Borisova2016,Cantalupo2019,FAB2019}.

\section{Results} \label{results} 
\subsection{Morphology and Emission}
\begin{figure*}[b!t]
    \centering
    \includegraphics[width=\textwidth]{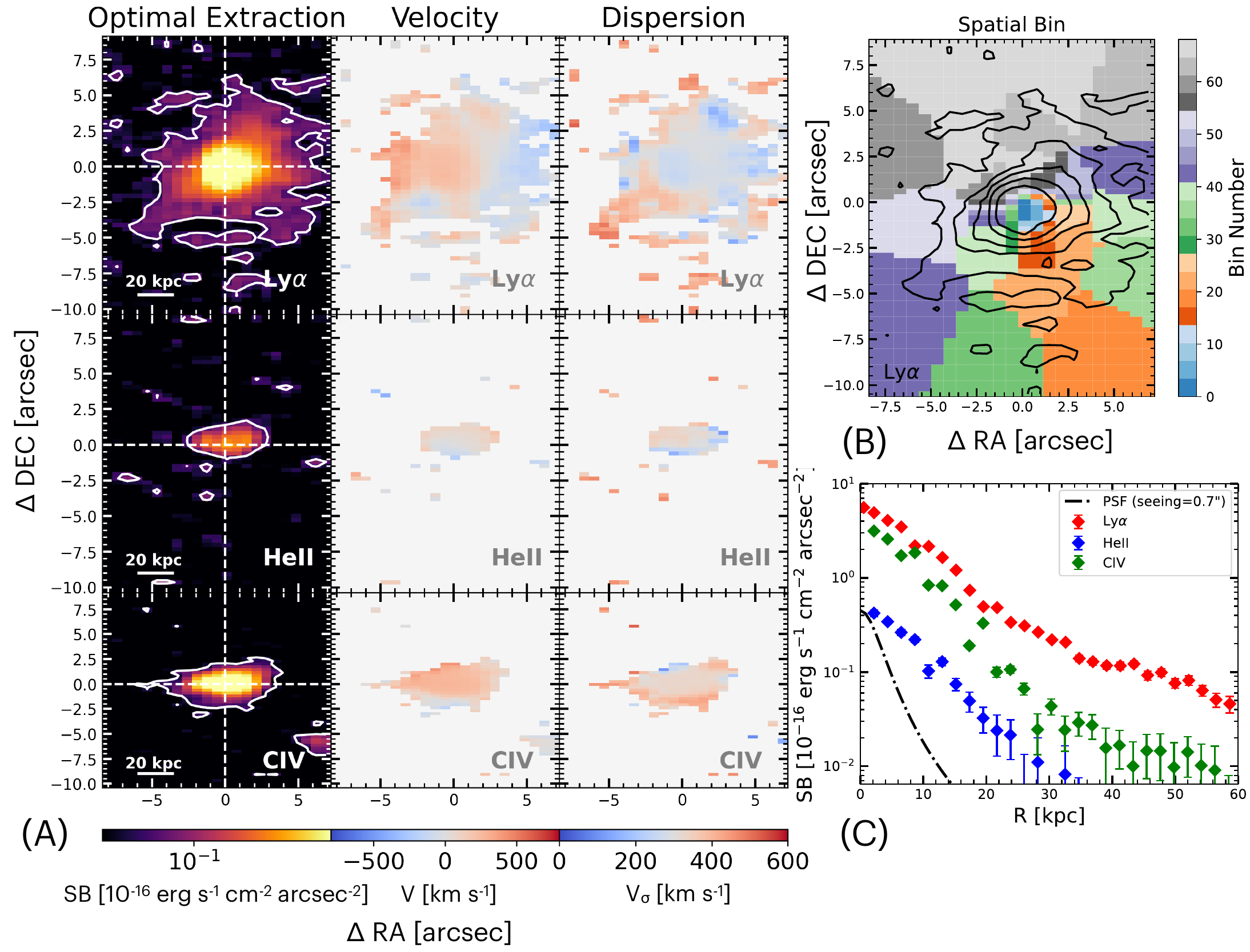}
    \caption{{\bf Panel A:} Optimally extracted images of the Ly$\alpha$, \heii, and \civ \ emissions. 
    White contours represent the 2$\sigma$ noise level.
    The three emission lines are detected on the scale of 122 kpc, 45 kpc, and 79 kpc, respectively. 
    {\it Middle:} The flux-weighted velocity maps. 
    {\it Right:} The flux-weighted dispersion map.
    {\bf Panel B:} The map of spatial bins coded by the color. 
    Black contours represent noise levels of [2$\sigma$, 5$\sigma$, 10$\sigma$, 20$\sigma$, 40$\sigma$] of the Ly$\alpha$ emission. 
    {\bf Panel C:} The SB profile of the Ly$\alpha$ (red), \heii \ (blue), and \civ \ (green) emissions compared with the PSF (black). 
    The Moffat profile is employed to model the PSF with the spatial resolution of $0.7''$. 
    The flux peak of the PSF is normalized to align with the \heii \ emission.}
    \label{data_map_spatial_bin}
\end{figure*}
The pseudo narrowband images of the Ly$\alpha$, \heii, and \civ \ emissions are shown in  Fig.~\ref{data_map_spatial_bin}.
The nebular emissions are enclosed in white contours which represent $S/N= 2$. 
By comparing the SB profile with the point spread function (PSF) shown in Fig.~\ref{data_map_spatial_bin}, we find that the Ly$\alpha$, \heii, and \civ \ emissions are all extended.
Within the 2$\sigma$ contours, the spatial extent of the three diffuse emissions is 122 kpc, 45 kpc, and 79 kpc, respectively. 


The integrated fluxes of the three diffuse emissions are $F_{\rm Ly\alpha}=5.30\pm 0.02\times 10^{-15}$ erg s$^{-1}$ cm$^{-2}$, 
$F_{\rm HeII}=2.79\pm 0.08 \times 10^{-16}$ erg s$^{-1}$ cm$^{-2}$, 
and $F_{\rm CIV}=2.02\pm 0.01\times 10^{-15}$ erg s$^{-1}$ cm$^{-2}$.
With the luminosity distance of $D_{\rm L}\approx 21868$ Mpc at $z_{\rm sys}$, these fluxes correspond to the luminosities of $L_{\rm Ly\alpha}=3.04\pm 0.02\times 10^{44}$ erg s$^{-1}$, $L_{\rm HeII}=1.60\pm 0.04\times 10^{43}$ erg s$^{-1}$, and $L_{\rm CIV}=11.53\pm 0.05\times 10^{43}$ erg s$^{-1}$, respectively.

\begin{figure*}[b!t]
    \centering
    \includegraphics[width=\textwidth]{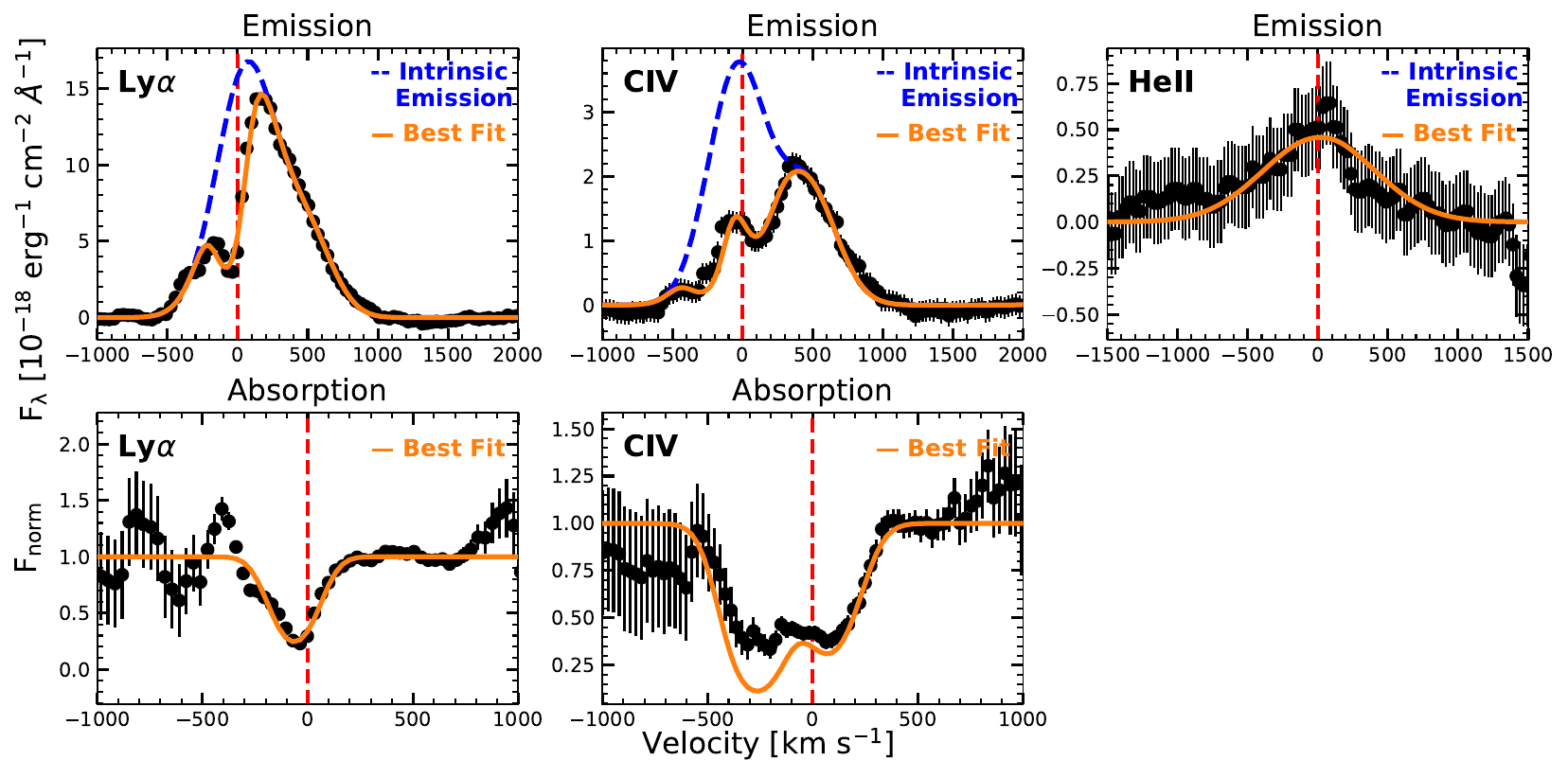}
    \caption{The spectra of the Ly$\alpha$, \heii, and \civ \ lines in bin-0. 
    We use the Gaussian profile and the voigt profile to fit the emission and the absorption, respectively.
    Blue dashed lines represent the emission component and the orange solid lines are the best fits (including the absorption). 
    For \civ \ absorption, the overshoot of the model to the data in the velocity range of $-500 \ {\rm km \ s^{-1}} \leq v \leq 0 \ {\rm km \ s^{-1}}$ is the consequence of reaching the smallest residual.
    Red dashed lines mark the zero point for each emission with the systemic redshift. 
    The results clearly demonstrate that our fittings are consistent with the observed spectra.}
    \label{bin0spec}
\end{figure*}
\subsection{Spectral Analysis}\label{spectra_analysis}
Here, we present the result of the spatially resolved spectra.
We use the Voronoi binning technique \citep{Cappellari2003} to spatially bin the IFS data. 
This technique adaptively bins the pixel of the input IFS data to reach a user-defined $S/N$ threshold. 
We set the $S/N$ threshold to $S/N=25$ 
estimated from the wavelength range of $4410 - 4450 \ {\rm \AA}$ which fully covers the Ly$\alpha$ emission. 
Note that this $S/N$ threshold is derived from the un-optimally-extracted image. 
By adopting the $S/N=25$, the Voronoi binning gives 69 spatial bins in total (Fig.~\ref{data_map_spatial_bin}).
The spectra of the Ly$\alpha$, \heii, and \civ \ emissions from the bin-0 are shown in Fig.~\ref{bin0spec} as an example. 
The \heii \ emission clearly shows one peak while both the Ly$\alpha$ and \civ \ emissions show two peaks with an enhanced redshifted peak. 
Since Ly$\alpha$ line exhibits different features from the \heii \ emission line, such features are not likely to be caused by the gas kinematics. 
Moreover, the scattering could also play a role as the resonant lines, Ly$\alpha$ and \civ, are more extended than \heii. 
Whereas, the resonant scattering is hard to reproduce the observed spectral features of Ly$\alpha$ and \civ \ since it requires the outflow with constant velocity \citep{chang2022} which is unnatural on the CGM scale.
Actually, recent works \citep{Wang2021,Kolwa2019} find Ly$\alpha$ nebulae around high-redshift radio galaxies associated with the \hi \ and the \civ \ absorption with line features similar to our results. 
Moreover, the absorption features are seen at the similar redshift for both Ly$\alpha$ and \civ \ lines. 
Thus, the absorption scenario is more favored.


\begin{figure*}[b!t]
    \centering
    \includegraphics[width=\textwidth]{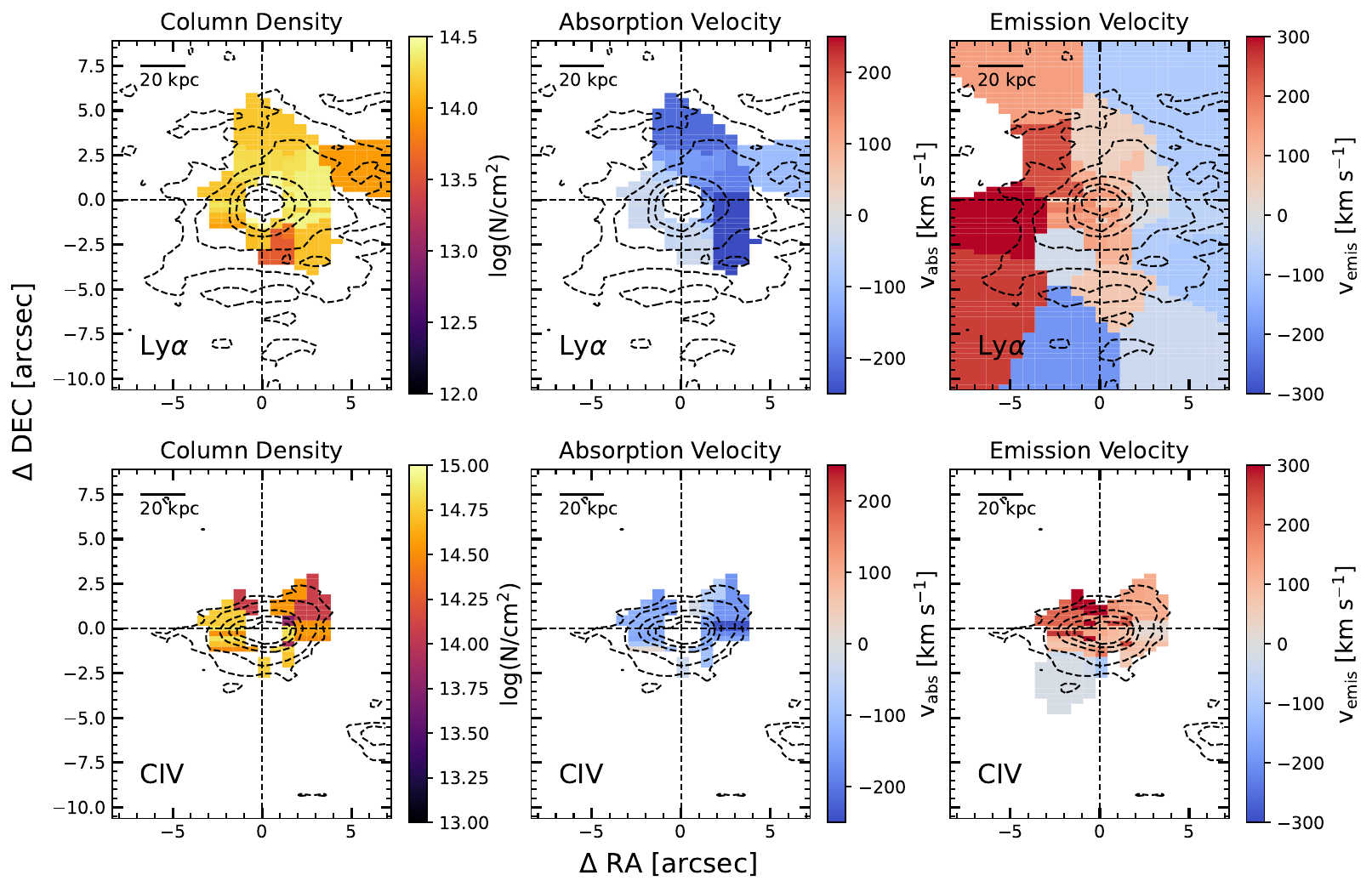}
    \caption{The parameter maps by fitting the spectra of the Ly$\alpha$ and the \civ. 
    The black contours denote the emission region with the noise level of [2$\sigma$, 5$\sigma$, 20$\sigma$, 40$\sigma$, 60$\sigma$]. 
    For Both emissions and absorption, the spatial bins with $S/N<3$ are removed. 
    The spatial bins within the central 10 kpc are also removed to avoid the influence of the PSF.}
    \label{paramap}
\end{figure*}

In addition, \cite{alexandroff2013} show that the Q1517+0055 is a type-II quasar which has no significant broad component. 
By fitting the spectra extracted from bin-0 where the flux peaks, we find that the line width (FWHM) of the Ly$\alpha$, \civ, and \heii \ emissions are $\rm FWHM_{Ly\alpha}=503\pm 17 \ km \ s^{-1}$, $\rm FWHM_{CIV}=482\pm 23 \ km \ s^{-1}$, and $\rm FWHM_{HeII}=459\pm 12 \ km \ s^{-1}$, respectively. 
Compared with the threshold of $\rm FWHM_{CIV}<2000 \ km \ s^{-1}$ used to select the type-II quasar \citep{alexandroff2013}, our measured line width is about four times smaller. 
This result suggests that the emissions are more likely to have no significant broad components.

The spectra extracted from 69 spatial bins are fitted. We follow the same procedure as \cite{Kolwa2019} and \cite{Wang2021} to fit spectra. 
We use the Gaussian function in the python package ASTROPY\footnote{https://www.astropy.org/} and Voigt function in the software LineTools\footnote{https://linetools.readthedocs.io/en/latest/\#} to model the emission and absorption.
The emergent emission is $F_{\lambda}=F_{\lambda,0}e^{-\tau_{\lambda}}$. The unabsorbed flux $F_{\lambda,0}$ is shown in Eq.~\ref{gaussian_emission}. 
\begin{equation}
    F_{\lambda,0}=Ae^{-0.5(\lambda-\lambda_{\rm emis})^{2}/\sigma^{2}}
    \label{gaussian_emission}
\end{equation}
where $A$ is the amplitude, $\lambda_{\rm emis}$ is the line center, and $\sigma$ is the line dispersion. 
The absorption is quantified by the optical depth, $\tau_{\lambda}$, which is shown in Eq.~\ref{depth_absorption}.
\begin{equation}
    \tau_\lambda=\frac{N \sqrt{\pi} e^2 f \lambda_0}{b m_{\mathrm{e}} c} H(a, u)
    \label{depth_absorption}
\end{equation}
where $N$ is the column density, $e$ is the electron charge, $m_{e}$ is the electron mass, $c$ is the speed of light, and $f$ is the oscillator strength. 
$H(a,u)$ is the Hjerting function where $a=\Gamma \lambda_{\rm abs}/4\pi cb$ and $u=(\lambda-\lambda_{\rm abs})/b\lambda_{\rm abs}$. 
$\Gamma$ is the Lorentzian width and $b$ is the Doppler parameter.

To fit the spectrum of Ly$\alpha$, \heii and \civ, we initialize $A$ as the peak flux of each spectrum, $\lambda_{\rm abs}$ as the observed wavelength at the systemic redshift, $z_{\rm sys}=2.6474\pm 0.0002$, and $\sigma$ as the width which corresponds to the result of primary fitting of \heii.
The column density, $N_{\rm HI}$, is constrained to a conservative range of $10^{12}$—$10^{20}$ cm$^{-2}$ by following the \cite{Kolwa2019}. 
The initial guess of the column density is taken as $N_{\rm HI}=10^{16}$ cm$^{-2}$ which is the mean of boundaries in log space. 
The Doppler parameter, $b$, is constrained in the range of 45—200 km s$^{-1}$ where the lower limit corresponds to the spectral resolution of FWHM=75 km s$^{-1}$. 
To fit the spectrum of \civ \ doublet, the emission flux ratio between the two lines is set to be $A_{1}/A_{2}=f_{1}/f_{2}\approx 2$ \citep{Kolwa2019,Wang2021} where the initial value of $A_{1}$ is set to be the peak flux in the spectrum. 
The \civ \ column density is constrained in the range of 10$^{13}$—10$^{16}$ cm$^{-2}$ by following \cite{Wang2021} with the 10$^{14.5}$ cm$^{-2}$ as the initial guess. 
The initial value of $\lambda_{\rm emis}$ and $\lambda_{\rm abs}$ for \civ \ are set in the same way as Ly$\alpha$. 
We present the initial value and range of fitting parameters in Tab.~\ref{lya_fittings}, \ref{civ_fittings}, and \ref{heii_fitting}. 
The best-fitting parameters are shown in Tab.~\ref{fitting_table} and \ref{fitting_table2}.

After pairing fitting parameters to corresponding spatial bins, we get parameter maps (Fig.~\ref{paramap}). For maps derived from fitting both the emission and the absorption, bins with the $(S/N)_{i}<3$ are removed where $(S/N)_{i}$ denotes the signal-to-noise ratio at bin $i$.

\subsection{CGM kinematics}
\begin{figure*}[b!t]
    \centering
    \includegraphics[width=\textwidth]{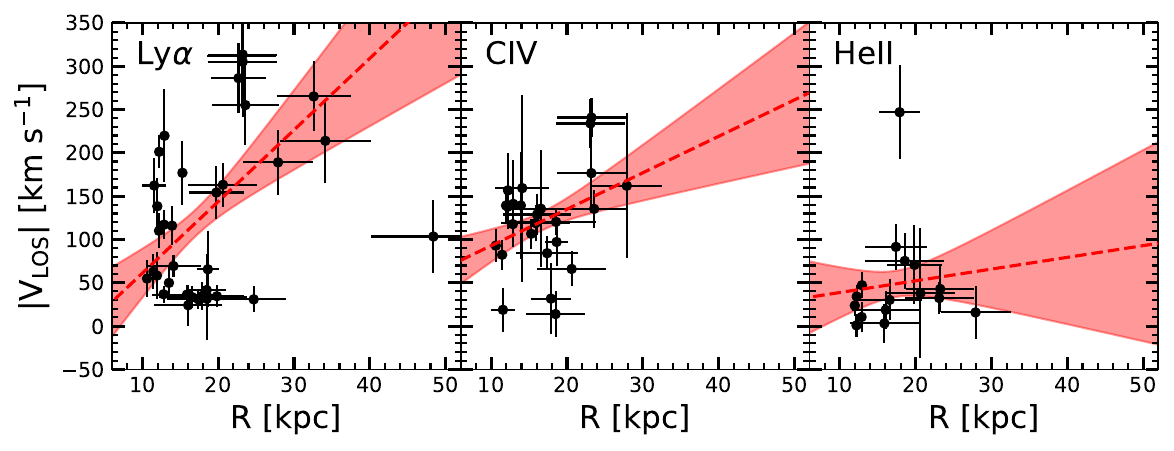}
    \caption{The absolute value of the line-of-sight velocity profile of the Ly$\alpha$ and \ \civ \ absorption, and the \heii \ emission. 
The black dots are from observations. 
The red dashed lines are the linear fitting to the black dots in the radius range of $10 \ {\rm kpc}\leq R \leq 30 \ {\rm kpc}$.  
The red shadow denotes the 1-$\sigma$ uncertainty of the fitting. 
The fitting indicates that the profile of the Ly$\alpha$ and \civ \ roughly have an increasing trend and the profile of the \heii \ is constant.}
    \label{vprofile}
\end{figure*}

We plot radial profiles of the absolute value of the line-of-sight velocity (referred as velocity profile in the following) yielded from the Ly$\alpha$ and \civ \ absorption and the \heii \ emission (Fig.~\ref{vprofile}).
The velocity is converted from the wavelength of the absorption/emission by 
\begin{equation}
    v=\frac{\lambda-\lambda_{r}(1+z_{\rm sys})}{\lambda_{r}(1+z_{\rm sys})}c
    \label{lambda2velocity}
\end{equation}
where $\lambda$ is the observed wavelength of the line, $\lambda_{r}$ is wavelength of the line in the rest-frame, $z_{\rm sys}=2.6474\pm 0.0002$ is the systemic redshift of Q1517+0055, and $c$ is the speed of the light.
The radius of the point is measured as the distance between the quasar and the spatial bin center. 
We remove the velocities from spatial bins with the radius of $R\leq 10 \ {\rm kpc}$ to avoid the influence of the PSF.

We fit the velocity profiles in the range of $10 \ {\rm kpc}< R \leq 30 \ {\rm kpc}$ with a linear model. 
The best-fitting slopes are shown in Tab.~\ref{vslope}. 
From the fitting results, we find that the profiles of Ly$\alpha$ and \civ \ have an increasing trend up to 30 kpc. 
For \heii, the profile is consistent with a constant model within the 1-$\sigma$ uncertainty. 
Moreover, The slopes yielded from the Ly$\alpha$ and \civ \ absorption are consistent with each other within 1-$\sigma$ uncertainty, which implies that the two absorptions might trace the same flow of gas. 
The increasing velocity profile could be a natural result of the gas inflow, which has been studied in observations \citep{zhang2022}, simulations \citep{Wangsen2021}, and semi-analytic models \citep{afruni2019,Lan2019}. 
In sec.~\ref{cgm_kinematics}, we present a detailed discussion.

\section{Discussion} \label{discussion}

\subsection{Photoionization Model} \label{powering_mechanism}
\begin{figure*}[b!t]
    \centering
    \includegraphics[width=\textwidth]{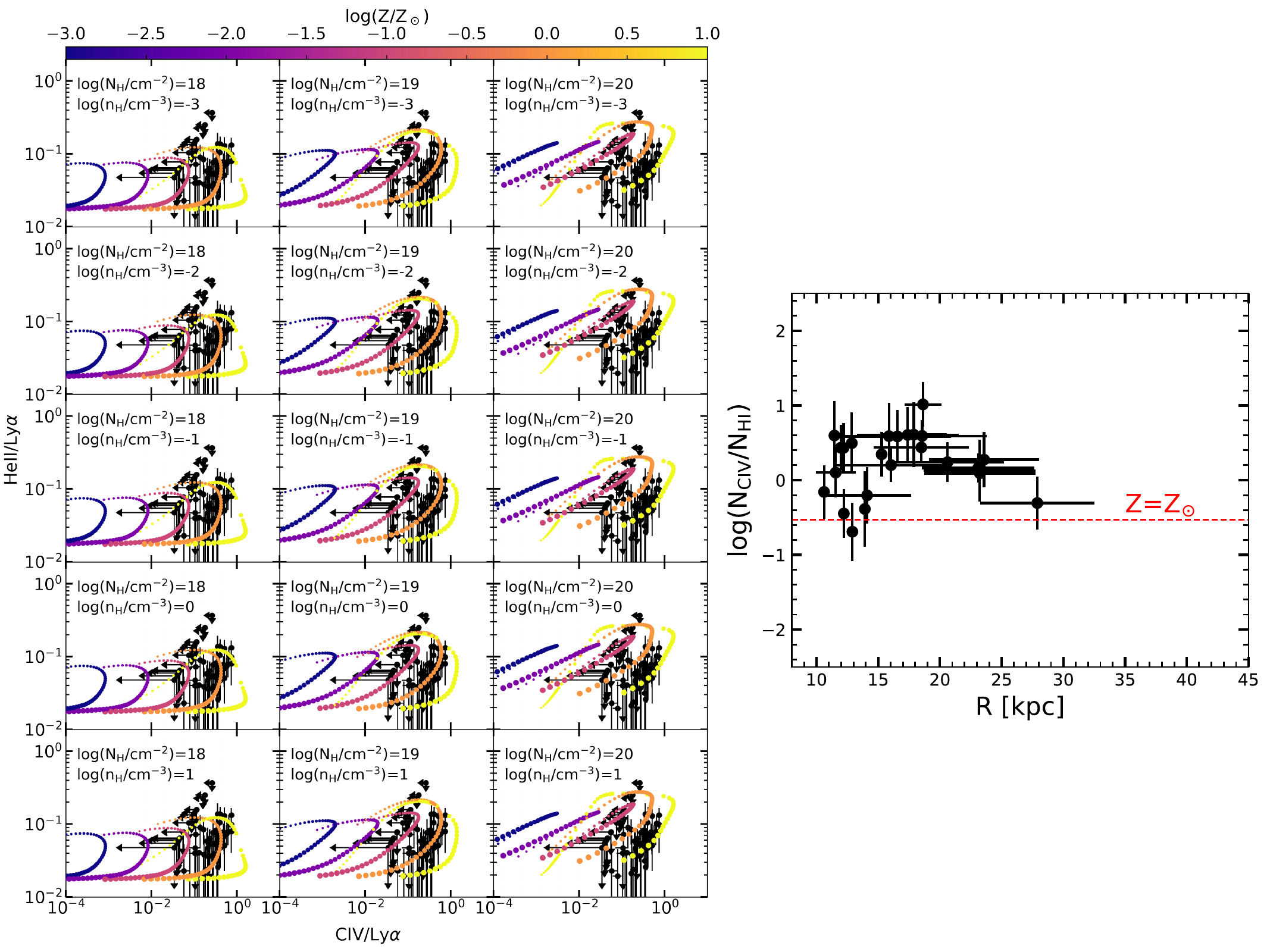}
    \caption{{\bf Left:} The comparison of observed line ratios (black) and modeled line ratios (colored). 
    The color and size of modeled points denote the metallicity and the ionization parameter, respectively.
    Black upper limits are from bins without \civ \ and  \heii \ emissions. 
    The consistency between observations and models indicates that diffuse emissions should be powered by photoionization with the metallicity of $Z\geq Z_{\odot}$.
    {\bf Right:} The column density ratio profile produced by fitting the absorption. 
    The red dashed line denotes the column density ratios with metallicity of $Z_{\odot}$ from CLOUDY simulations. 
    This figure shows that the cool gas traced by the absorption has the metallicity of $\geq Z_{\odot}$ out to 30 kpc.}
    \label{line_column_ratio}
\end{figure*}

The powering mechanism of these diffuse emissions could be inferred from the line ratio diagnostics.
We calculate the line ratio based on the fitted emission line (blue dashed lines in Fig.~\ref{bin0spec}) in each spatial bin (Fig.~\ref{data_map_spatial_bin}).
The velocity range used to extract the fluxes is set to  $-1000 \ {\rm km \ s^{-1}}\leq \Delta v \leq 1000 \ {\rm km \ s^{-1}}$ which fully covers the emission lines.
Then, the CLOUDY simulations \citep{Ferland2017} are used as follows.  
We apply the built-in AGN template as the incident radiation. 
The UV slope of the AGN template is adopted as $\alpha_{\rm uv}=-2$ to keep consistent with observations.
As for model parameters, we follow previous work \citep{Cai2017,FAB2015} to set the ionization parameter, $log(U)$, in the range of $-3-0$ with a step of 0.06, the metallicity, $Z$, in the set of [$10^{-3}$, $10^{-2}$, $10^{-1}$, $1$, $10$]$Z_{\odot}$, the column density, $log(N_{\rm H}/{\rm cm^{-2}})$, in the range of $18-20$ with a step of $1$, the hydrogen number density, $log(n_{\rm H}/{\rm cm^{-3}})$, in the range of $-3-1$ with a step of $1$. 
Both the recombination radiation from the CGM and the resonant scattering of AGN's photons are included in the model. 
We show both the photoionization model (color dots) and observations (black dots) in Fig.~\ref{line_column_ratio}, Left.
Because of the high ratio between \civ \ and Ly$\alpha$ (\civ/Ly$\alpha \approx0.28$), the observed line ratio could be consistent with models of $Z\geq Z_{\odot}$. 
Considering that no significant broad component contaminates the emission (see Sec.\ref{spectra_analysis}), 
this consistency between the observed line ratios and the solar-metallicity model indicates that the cool gas in the CGM is metal-enriched.

Actually, under the photoionization scenario, the absorption also yields high metallicity.
Fig.~\ref{line_column_ratio}, Right presents the observed column density ratios profile of the \hi \ and \civ \ (black dots) versus the modeled column density ratios by CLOUDY simulations (red dashed lines). 
The observed column density ratio profile indicates metallicity of $Z\geq Z_{\odot}$ out to 30 kpc.

In fact, the metal-enriched CGM has been prevalently revealed by recent observations on the diffuse emission \citep{Guo2021,Fossati2021,Marques-Chaves2019,Kolwa2019} and absorption \citep{Prochaska2009,Prochaska2014}. 
Both \cite{Marques-Chaves2019} and \cite{Kolwa2019} reveal that the CGM could reach solar metallicity or beyond though observing the Ly$\alpha$ and \civ \ nebulae. 
Our observations are consistent with these results, which indicates that the CGM could be highly metal-enriched at the cosmic noon.

\subsection{Outflow scenario} \label{outflow_scenario}

The previous section reveals that the CGM of Q1517+0055 is enriched to about solar metallicity. 
Such high metallicity could be due to the AGN feedback. 
Under this scenario, the outflowing gas should be accelerated out to roughly 30 kpc to match our observations. 
In terms of the theory, 
although some hydrodynamic simulations demonstrate that the outflow could be accelerating on the scale of 10 kpc \citep{Mitchell2020}, most simulations of the accelerating outflow only focus on the scale of $\leq 1$ kpc \citep{Tanner2022}. 
In terms of the observations, the accelerating outflows are only observed on the scale of hundreds of parsec to 1 kpc \citep{Santoro2020,He2022}. 
Although \cite{David2019} directly imaging the AGN feedback on the physical scale of 100 kpc at $z\approx 0.5$, previous observations reveal that the outflow is mostly decelerating on the large scale \citep{Kacprzak2019,Ng2019,David2019}.

Besides, we also calculate the coupling efficiency ($f_{c}$) which is the ratio between the outflow energy rate ($\dot{E}_{\rm out}$) and the active galactic nucleus (AGN) luminosity ($L_{\rm AGN}$). 
This value describes the efficiency of the AGN luminosity coupling to the interstellar medium (ISM) or CGM. 
By fitting the spectral energy distribution (SED) of Q1517+0055 with \textsc{cigale} \citep{Boquien2019}, we derive the AGN luminosity to be $L_{\rm AGN}=8.0\times 10^{45}$ erg s$^{-1}$. 
For $\dot{E}_{\rm out}$, we follow previous works \citep{Harrison2012,Harrison2014} to calculate the upper limit and the lower limit. 

For the upper limit, we adopt the energy-conserved  model that assumes an inefficient cooling process. 
This model gives \citep{Harrison2012,Harrison2014,Cai2017}
\begin{equation}
    \dot{E}_{\rm out}=1.5\times 10^{46} R^{2}_{\rm out}v^{3}_{\rm out}n_{e} \ {\rm erg \ s^{-1}}
    \label{E_out_up}
\end{equation}
where $R_{\rm out}$ is the outflow radii in the unit of 10 kpc, $v_{\rm out}$ is the outflow velocity in the unit of 1000 km s$^{-1}$, and $n_{e}$ is the electron number density in the unit of 0.5 cm$^{-3}$. 
By adopting our observational results, $R_{\rm out}=$39 kpc (half the spatial extent of \civ \ emission), $v_{\rm out}=320$ km s$^{-1}$ (the max velocity yielded from the \hi \ absorption), and assuming electron density of $n_{e}=1.5\ {\rm cm^{-3}}$, we calculate the upper limit of $\dot{E}_{\rm out}$ to be $2.3\times 10^{46}$ erg s$^{-1}$.

For the lower limit, we adopt the model which assumes the outflow extending from 0 kpc to where we see the extended emission. 
The $\dot{E}_{\rm out}$ is given by \citep{Harrison2012,Harrison2014,Cano2012,Greene2012}
\begin{equation}
    \dot{E}_{\rm out}=\frac{1}{2}\dot{M}_{\rm out}(v_{\rm out}^{2}+3\sigma_{v}^{2})
    \label{outflow_energy_rate}
\end{equation}
where $\dot{M}_{\rm out}$ is the outflow mass rate and $\sigma_{v}$ is the velocity dispersion. 
For $\dot{M}_{\rm out}$, we have $\dot{M}_{\rm out}=M_{\rm out}v_{\rm out}/R_{\rm out}$ where $M_{\rm out}$ is the outflow mass. 
Following \cite{zhang2022}, we derive the outflow mass to be $\dot{E}_{\rm out}=1.4\times 10^{45}$ erg s$^{-1}$. 
Then, the outflow energy rate is adopted as the mean of the upper limit and lower limit in the log space \citep{Harrison2012,Harrison2014}, which is $\dot{E}_{\rm out}=5.7\times 10^{45}$ erg s$^{-1}$. 

\begin{figure}[b!t]
    \centering
    \includegraphics[width=\columnwidth]{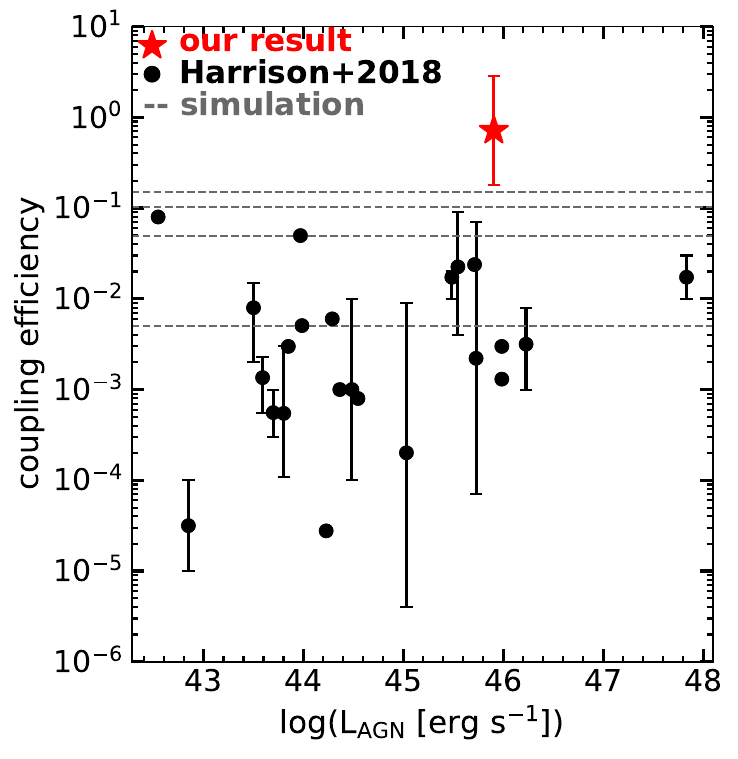}
    \caption{A comparison of the coupling efficiency calculated from our observations, assuming an AGN outflow model (red star) and other observations (black dots) and simulations (dashed lines) \citep{Harrison2018}. 
    The error bar of our result represents the upper and lower limit. 
    The coupling efficiency of $f_{c}\approx0.7$ is higher than those from simulations and previous observations.}
    \label{f_c}
\end{figure}

Given the outflow energy rate and AGN luminosity, we calculate the coupling efficiency to be $f_{c}=0.7$. 
The results are shown in Fig.~\ref{f_c}. 
Compared with previous observations and simulations \citep{Harrison2018}, the lower limit of $f_{c}$ for our observations is slightly higher than the largest $f_{c}$ predicted by simulations. 
The mean $f_{c}$ for our observations is about five times larger than this largest value. 
Based on the discussion and estimations above, we note that the extremely powerful AGN feedback might be an interpretation of our observations. 
Such feedback should be able to accelerate the gas on the scale out to 30 kpc and have the $f_{c}$ reaching $\approx 0.7$.

\subsection{Inflow scenario} \label{cgm_kinematics}
\begin{figure*}[b!t]
    \centering
    \includegraphics[width=\textwidth]{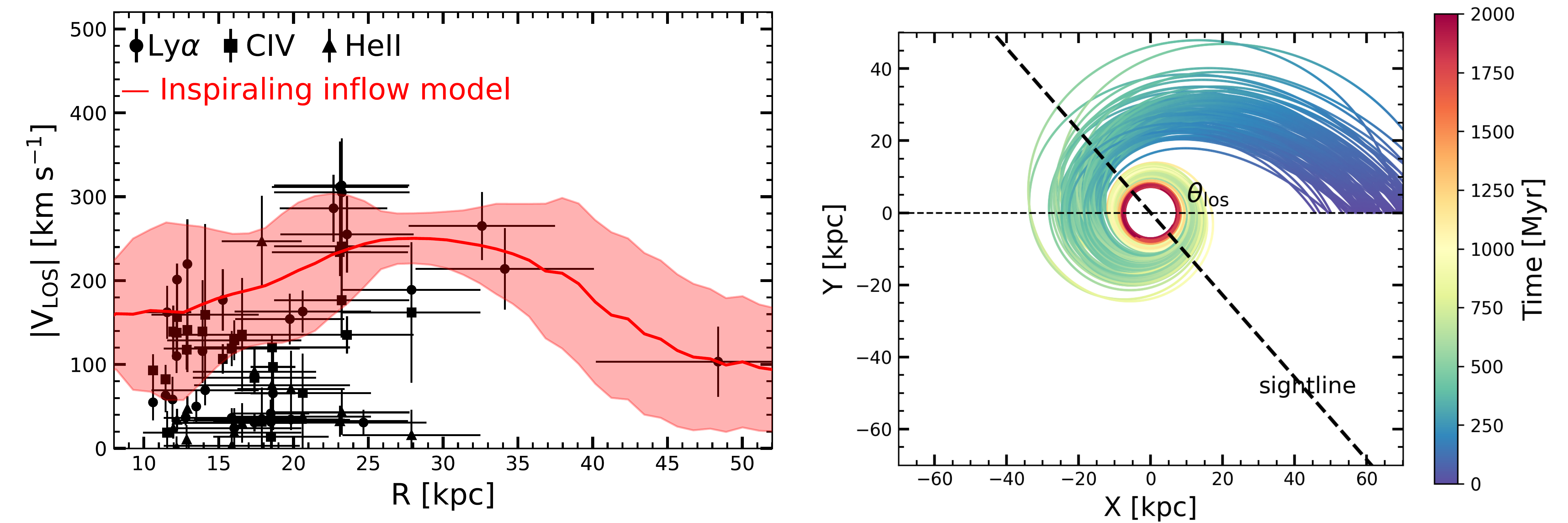}
    \caption{{\bf Left:} The line-of-sight velocity profile gotten by fitting spectra. 
    Here, we only show the absolute value of the velocity.
    The errorbar of the velocity is from the fitting while the errorbar of the radius represents the 1$\sigma$ scatter of the spatial bin. 
    The red line is the simulated line-of-sight velocity profile from our model. 
    The red shadow represents the $2\sigma$ scatter. 
    This figure shows that the velocity profile increases and then drops down from the inside out, which can be reproduced by the inflow model.
    {\bf Right:} The trajectories of the inflowing cloud. 
    The color codes the inflowing time. 
    This set of trajectories is yielded by ranging the fitting parameters in the $2\sigma$ range. 
    The dashed line marks the direction of the sightline. 
    $\theta_{\rm los}$ is the angle describing the direction of the sightline. 
    This result means that our inflow model can produce the inspiraling inflow seen in both observations \citep{zhang2022} and simulations \citep{Wangsen2021}.}
    \label{vmodel_track}
\end{figure*}

In fact, the observed increasing velocity profile can be produced by the gas inflow.
Such kinematics is not proposed for the first time. 
The CGM kinematics has been studied by observations \citep{Weidinger2004,Weidinger2005,fab2018,zhang2022}, simulations \citep{Wangsen2021}, and semi-analytical models \citep{afruni2019,afruni2022,Lan2019}.
\cite{zhang2022} observe the Ly$\alpha$ nebula with the diffuse \heii \ and \civ \ emissions around a type-II quasar at $z=2.3$, revealing an increasing velocity profile in the nebula which represents the metal-enriched inflow. 
\cite{Wangsen2021} found similar infalling CGM gas behavior in the IllustrisTNG simulations. 
The cool circumgalactic gas speeds up as it inflows from larger radii. 
Both \cite{afruni2019} and \citet{Lan2019} construct a semi-analytic model showing the cool gas cloud could accelerate in the CGM.

To understand the physics of the observed kinematics, an inspiraling inflow model with the gas angular momentum included is constructed in this work. 

\cite{afruni2019} and \cite{Lan2019} construct the semi-analytic model describing the radial motion of the cool gas inflow in the CGM.
Their inflow model shows that the cool gas motion is controlled by the gravitational force of the dark matter halo and the drag force of the hot gas.
The radial motion is described by Eq.~\ref{radial_motion} (Eq.~12 in \cite{afruni2019}),
\begin{equation}
    \frac{dv_{r}}{dt}=\frac{GM(r)}{r^{2}}-\frac{\pi r_{\rm cool}^{2}\rho_{\rm hot}(r)v_{ r}^{2}}{m_{\rm cool}}
    \label{radial_motion}
\end{equation}
where $G$ is the gravitational constant, $v_{r}$ is the velocity of the cool gas relative to the hot gas, $M(r)$ is the mass within the radius of $r$ following the Navarro Frenk White (NFW) profile \citep{afruni2019}, $r_{\rm cool}$ is the radius of the cool gas cloud assuming that the cloud is spherical, $\rho_{\rm hot}(r)$ is the density of the hot gas, and $m_{\rm cool}$ is the mass of the cool gas cloud. 
For the hot gas density, we have $\rho_{\rm hot}=f_{\rm hot}\rho_{\rm bar}$ where $\rho_{\rm bar}$ is the baryon density and $f_{\rm hot}$ is the hot gas fraction. 
Since there is almost no observational constraint for the hot gas fraction at high-$z$, $f_{\rm cor}$ is set to be a free parameter here.
Details about how $\rho_{\rm bar}$ and $r_{\rm cool}$ are derived are shown in \citep{afruni2019}.
On the right side of Eq.~\ref{radial_motion}, the first term represents the gravity of the halo, and the second term represents the drag force of the hot gas.

Since the tangential motion of the cool gas cloud is also important, we include it in the model.
The equation of the drag force should be modified because it is determined by the relative velocity. 
The Eq.~\ref{radial_motion} is changed to 
\begin{equation}
    \frac{dv_{r}}{dt}=\frac{GM(r)}{r^{2}}-\frac{\pi r_{\rm cool}^{2}\rho_{\rm hot}(r)v^{\prime}v_{r}}{m_{\rm cool}}-\frac{v_{t}^{2}}{r}-\frac{dm_{\rm cool}}{dt}\frac{v_{r}}{m_{\rm cool}}
    \label{radial_motion2}
\end{equation}
where $v_{t}$ is the tangential velocity and $v^{\prime}=\sqrt{(v_{t}-v_{\rm hot})^{2}+v_{r}^{2}}$ is the velocity of the cool gas relative to the hot gas, 
$v_{\rm hot}$ is the rotation velocity of the hot gas.
On the right side of Eq.~\ref{radial_motion2}, the first term is the gravitational force from the halo, and the second term is the projection of the drag force in the radial direction.
The newly added two terms ($v_{t}^{2}/r$, $\frac{dm_{\rm cool}}{dt}\frac{v_{r}}{m_{\rm cool}}$) represent the centripetal force due to the tangential motion and the mass change of the cool gas cloud, respectively.
The equation describing the tangential motion is
\begin{equation}
    \frac{dv_{t}}{dt}=-\frac{\pi r^{2}\rho_{\rm hot}(r)v^{\prime}(v_{t}-v_{\rm hot})}{m_{\rm cool}}-\frac{dm_{\rm cool}}{dt}\frac{v_{t}}{m_{\rm cool}}-\frac{v_{r}v_{t}}{r}
    \label{tangential_motion}
\end{equation}
where, on the right side, the first term is the projection of the drag force in the tangential direction, the second term represents the mass change, and the third term represents the influence of the tangential motion.
The equation describing the mass change of the cool gas is
\begin{equation}
    \frac{dm_{\rm cool}}{dt}=-\alpha m_{\rm cool}
\end{equation}
where $\alpha$ is the evaporation rate describing the mass loss rate of the cool gas cloud \citep{afruni2019}.

The dark matter halo mass is needed for the inflow model. 
By fitting the spectral energy distribution (SED) of Q1517+0055, we derive the stellar mass of the host galaxy to be ${\rm log}(M_{\star}/M_{\odot})=10.5^{+0.1}_{-0.1}$.
Through the stellar-mass-to-halo-mass relation of \cite{lu2014}, we get the halo mass of Q1517+0055 to be ${\rm log}(M_{\rm h}/M_{\odot})=12.1^{+0.3}_{-0.1}$.

\begin{figure*}[b!t]
    \centering
    \includegraphics[width=\textwidth]{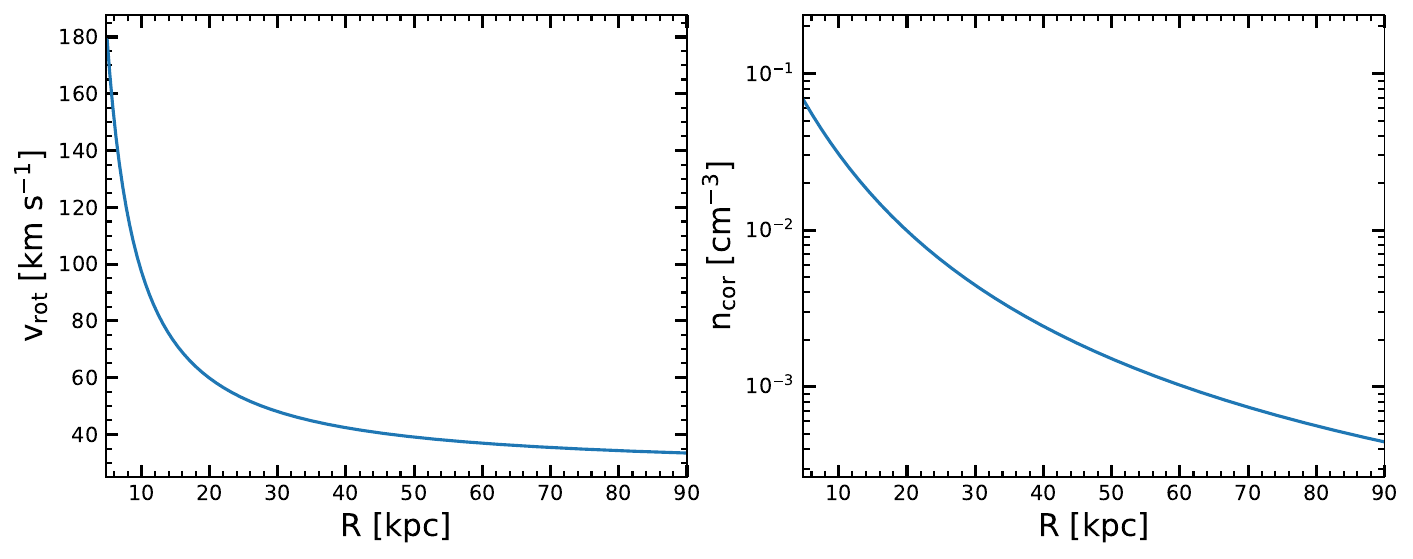}
    \caption{{\bf Left:} The rotation profile of the hot gas in the CGM. 
    We construct this rotation profile by following the result of \cite{Danovich2015} which study the angular momentum of the CGM gas for galaxies at $z=1\sim3$. {\bf Right:} The density profile of the hot gas. This density profile is constructed by following the Eq.~6 in \cite{afruni2019}.}
    \label{hot_gas_properties}
\end{figure*}

To model the hot gas rotation, we use the result of \cite{Danovich2015} which studies the angular momentum of CGM at $z=1.5\sim 4$ in cosmological simulations.
In their simulations, the systems have the halo mass of 10$^{11.4\sim12.2}$ $M_{\odot}$ at $z\approx2$, which are consistent with our observations. 
To compare with the observed line-of-sight velocity, we further introduce the angle of sightline, $\theta_{\rm los}$, which is used for the projection (Fig.~\ref{vmodel_track}, right). 
The rotation and density profile of the hot corona are shown in Fig.~\ref{hot_gas_properties}.


Based on the description above, the inspiraling inflow model has seven free parameters. 
They are the initial infalling radius, $r_{\rm init}$, the initial radial velocity of the cloud, $v_{\rm r, init}$, the initial tangential velocity of the cloud, $v_{\rm init, t}$, the initial mass of the cloud, $m_{\rm cool, init}$, the mass evaporation rate, $\alpha$, the hot corona fraction, $f_{\rm cor}$, and the angle of the sightline, $\theta_{\rm los}$.
We then fit the observed line-of-sight velocity profile (Fig.~\ref{vmodel_track}) using a Markov chain Monte Carlo (MCMC) analysis.
Considering that the velocity profile of Ly$\alpha$ \ not only exhibits a similar trend with \civ \ but also is observed at the largest radius, we only use the velocity profile of Ly$\alpha$ \ here. 
Eq.~\ref{likelihood} denotes the logarithm of the likelihood used for performing the MCMC. 
\begin{equation}
\ln \mathcal{L}=-\frac{1}{2}\Sigma_{i}\frac{(v_{\rm obs}(r_{i})-v_{\rm model}(r_{i}))^{2}}{\sigma^{2}_{\rm obs}(r_{i})}
\label{likelihood}
\end{equation}
The prior of free parameters are shown in Tab.~\ref{parameter_boundary}.
The fitting results are shown in Fig.~\ref{vmodel_track} (left).
The reduced $\chi^{2}$ is $\chi_{r}^{2}\approx 1.3$ which indicates the modified model is consistent with our observations.

\begin{figure*}[b!t]
    \centering
    \includegraphics[width=\textwidth]{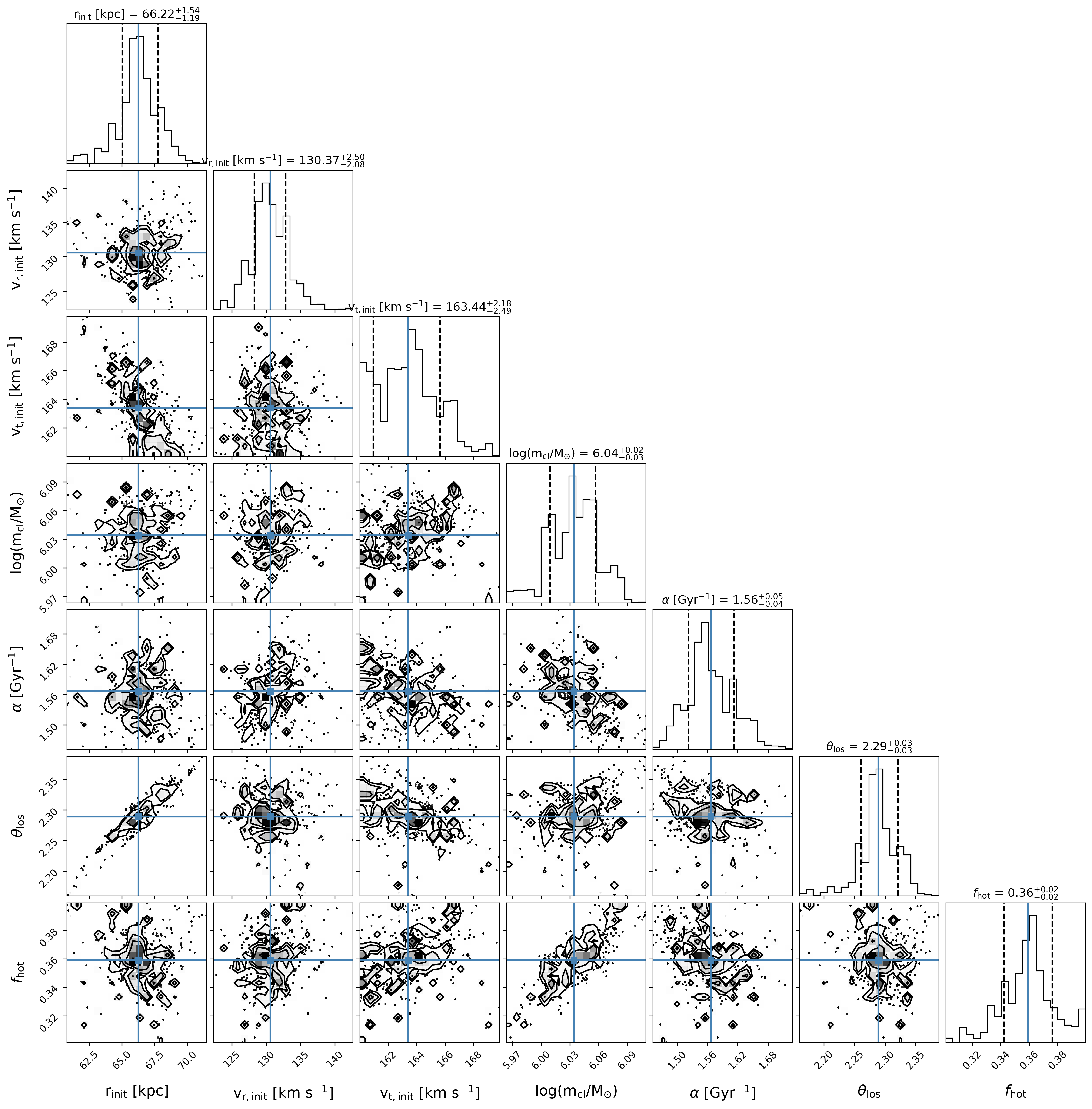}
    \caption{Corner plot of the MCMC results for the fitting parameters. 
    The one and two-dimensional posterior probabilities for the seven free parameters are shown. 
    We initialize the twenty-five Markov chains and iterate each chain for twenty-thousand steps. 
    The first fifty steps are discarded for burn-in. 
    After the iteration, the results converge.}
    \label{mcmc}
\end{figure*}

The corner plot of the best-fitting parameters are shown in Fig.~\ref{mcmc}. 
The best-fit initial radius, $r_{\rm init}=66.22^{+1.54}_{-1.19}$ kpc, is within the halo virial radius of 87 kpc. 
We note that the initial radius is limited by observations. 
The velocities are only observed up to 50 kpc.
The best-fit initial radial velocity and tangential velocity of the cool cloud are $v_{\rm r, init}=130.37^{+2.5}_{-2.08}$ km s$^{-1}$ and $v_{\rm t, init}=163.44^{+2.18}_{-2.49}$ km s$^{-1}$, respectively. 
\cite{Wangsen2021} show that the cool CGM gas could gain angular momentum from the large-scale environment through galaxy fly-by interactions which lead the gas velocity to be $0-200$ km s$^{-1}$. 
In fact, recent observations have revealed the CGM inflow with a radial velocity of $\sim 300$ km s$^{-1}$ at $z\approx2$ \citep{Fu2021}. 
The best-fitting radial velocity and tangential velocity are comparable to both simulations and observations. 
For the evaporation rate, our best-fit value is $\alpha=1.56^{+0.05}_{-0.04}$ Gyr$^{-1}$ which is similar to the low-$z$ results \citep{afruni2019}.  
For the mass of the cool cloud, we find that our best-fitting value is ${\rm log}(m_{\rm cl}/M_{\odot})=6.04^{+0.02}_{-0.03}$. 
This mass is two orders of magnitude larger than the cool cloud mass estimated by \cite{afruni2019} but roughly consistent with the cool cloud mass yielded by \cite{afruni2022}. 
For the hot corona fraction, we derive $f_{\rm hot}=0.36^{+0.02}_{-0.02}$ which is consistent with cosmological simulations \citep{Machado2018}. 
Note that, from Fig.~\ref{mcmc}, the cloud mass degenerates with the evaporation rate and the hot corona fraction ($f_{\rm hot}$). 
We, thus, prefer not to have strong claims for these quantities.

From the inflow model, the velocity profile can be explained by the competition between the halo gravity and the drag force. 
The projection effect also plays a role here.
When the cool gas cloud starts to fall into the galaxy, the initial velocity is small. 
The cloud is firstly accelerated due to the halo gravity at this stage.
Since the drag force exerted by the hot gas is proportional to $v'^{2}$ (Eq.~\ref{radial_motion}), at the stage when the velocity of the cloud is large enough near the central region, the drag force of the hot gas and the loss of cool cloud mass begin to take control leading to the deceleration. 
In addition, the evaporation of the gas is also an important effect influencing the cloud motion. 
Moreover, the inspiraling trajectory of the gas inflow seen in both observations \citep{zhang2022} and simulations \citep{Wangsen2021,stewart2017} is the natural consequence after considering the tangential motion of both the cool gas and the hot gas (Fig.~\ref{vmodel_track}, right). 

We should note that our inflow model is simplified which only includes one gas cloud into consideration. 
In fact, the inflow should be a ``continuous flow'' with multiple clouds. 
These clouds could have different kinematics due to the gas shock. 
Since the detailed study of the gas shock is beyond this work, we present a brief discussion here. 

To induce the shock, the gas velocity should exceed the local sound speed ($c_{s}$) which is \citep{Yun2019}
\begin{equation}
    c_{s}=\sqrt{\frac{\gamma k_{\rm B}T_{\rm gas}}{\mu m_{\rm p}}}
    \label{sound_speed}
\end{equation}
where $T_{\rm gas}$ is the gas temperature, $k_{\rm B}$ is the Boltzmann constant, $\gamma=5/3$ is the adiabatic index for the monatomic gas, and $\mu m_{\rm p}$ is the average particle mass. 
$m_{\rm p}$ is the proton mass. 
For CGM, we have $\mu m_{\rm p}\approx 0.6m_{\rm p}$ \citep{Yun2019,Gritton2017}. 
In our case, the cloud is moving in the diffuse hot corona. 
Since the hot corona is virialized \citep{afruni2019}, the temperature of the hot corona is 
\begin{equation}
    T_{\rm gas}=\frac{\mu m_{\rm p}GM_{\rm h}}{2k_{\rm B}r_{\rm vir}}
    \label{T_vir}
\end{equation}
where $M_{\rm h}$ is the halo mass and $r_{\rm vir}$ is the virial radius of the halo. 
Given $M_{\rm h}=10^{12.1} \ M_{\odot}$ and $r_{\rm vir}=87$ kpc, we have $T_{\rm gas}=2.3\times 10^{6} \ {\rm K}$. 
From Eq.~\ref{sound_speed}, we have $c_{s}\approx228$ km s$^{-1}$. 
This local sound speed is roughly consistent with the largest value of the observed velocities, which indicates that the cloud is hard to induce the shock.
Nevertheless, considering that the observed velocity is only a projection of the gas velocity, the gas velocity should be larger than the observed velocity. 
Thus, for the ``continuous flow'' of the gas cloud, a small fraction of the cloud could be influenced by the shock. 

For most of the clouds, since they do not induce shock, they should have similar kinematics with each other (Fig.~\ref{vmodel_track}). 
For the small fraction of clouds influenced by the shock, we present the qualitative discussion.

On the one hand, the gas shock could slow down the inflowing gas by dissipating the kinetic energy. 
This will then make the velocity profile in the radius of $10 \ {\rm kpc}\leq r \leq 30 \ {\rm kpc}$ steeper. 
On the other hand, the drag force is proportional to the velocity squared. 
Since the cloud is slowed down by the shock, the drag force could be reduced.
This result will in turn flatten the velocity profile in the inner region. 
Combining these two effects, the velocity profile (Fig.~\ref{vmodel_track}, left) might not change too much after considering the shock. 
Moreover, the shock could also induce gas heating by converting the kinetic energy to the thermal energy of the gas, which might accelerate the evaporation of the cool cloud.
Nevertheless, for Q1517+0055, the gas is metal-enriched to about solar metallicity. 
The heated gas could also be efficiently cooled down through the emission from the highly ionized metal atoms such as the \civ \ emission. 
To reach a more solid conclusion, detailed hydrodynamical simulations and inflow models are needed to figure out the impact of the shock.



\subsection{CGM Gas Recycling} \label{gas_recycling}

\begin{figure*}[b!t]
    \centering
    \includegraphics[width=0.7\textwidth]{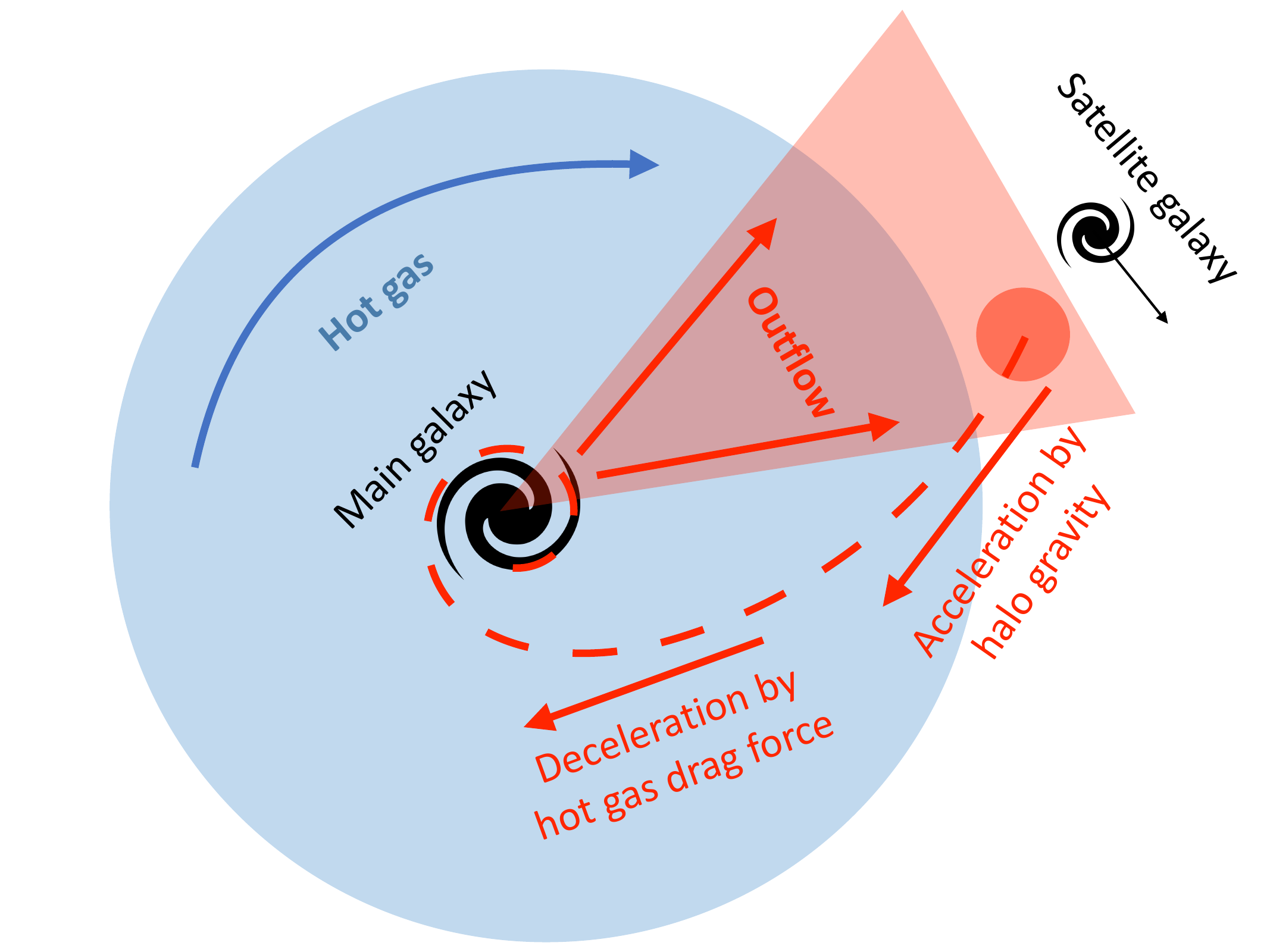}
    \caption{The physical picture of the gas recycling.
    The main galaxy first ejects the metal-enriched materials to the CGM. 
    After cooling, the metal-enriched gas cools and begins to fall back to the main galaxy.
    Under the influence of the satellite galaxy and the hot gas, the falling material forms the inspiraling stream. 
    The competition between the gravity of the halo and the drag force of the hot gas lead to the cool gas accelerating at first and then decelerating.}
    \label{gasrecyling}
\end{figure*}

Discussions above show that the metal-enriched cool gas is more likely to inflow to Q1517+0055. 
Such cool gas is undergoing the gas recycling process.
By combining the results of \cite{zhang2022} and this work, we show a possible physical picture in Fig.~\ref{gasrecyling}.
At the beginning, the metal-enriched gas born in the star-forming region of the galaxy is ejected to the CGM through feedback.
After cooling through the UV line emission, the enriched gas starts to fall back to the galaxy given that thermal pressure cannot support the gravity of the halo. 
Meanwhile, satellite galaxies fly by to further bring angular momentum to the CGM gas leading to the inspiraling form of the inflow \citep{Wangsen2021}.



We demonstrate that gas recycling could be a non-negligible process in the metal-enriched CGM around high-$z$ quasars. 
To fully reveal the details of gas recycling, more observations on the CGM diffuse emissions are needed.

\section{Conclusions} \label{conclusion}

In this paper, we present the discovery of a Ly$\alpha$ nebulae around the type-II quasar, Q1517+0055.
Our main findings are summarized as follows:
\begin{itemize}
    \item Above the $2\sigma$ surface brightness limit, we reveal the diffuse emissions of Ly$\alpha$, \heii, and \civ \ on the physical scale of 122 kpc, 45 kpc, and 79 kpc, respectively (Fig.~\ref{data_map_spatial_bin}, left).
    The total Ly$\alpha$ luminosity is $L_{\rm Ly\alpha}= 3.04\pm 0.02 \times 10^{44} \ {\rm erg \ s^{-1}}$.
    From the spectra, we find that the Ly$\alpha$ and \civ \ emissions exhibit double peaks which are caused by the associated Ly$\alpha$ \ and \civ \ absorption (Fig.~\ref{bin0spec}).
    \item By fitting the absorption, we find that the velocity of the absorption has an increasing trend (Fig.~\ref{vprofile}).  
    We construct an inspiraling inflow model which explains the velocity profile as the consequence of the competition between the halo gravity and the drag force of the hot gas. 
    The model shows that gas evaporation is also an important effect in determining the cloud motion. 
    \item From the line ratio diagnostic, we find that the observed line ratios between the diffuse emissions are ${\rm HeII/Ly\alpha}\approx 0.08$ and ${\rm CIV/Ly\alpha}\approx 0.28$, consistent with the photoionization scenario with recombination and photon pumping included (Fig.~\ref{line_column_ratio}, Left). 
    By comparing ratios of both the emission and column density, we find the cool gas in the CGM is metal-enriched with the metallicity of $Z\geq Z_{\odot}$.
    \item Comparing with the scenario of a powerful outflow, the inflow seems to be a more natural interpretation. 
    Gas recycling could be a non-negligible process for high-$z$ quasars. 
    More observations on the CGM diffuse emission are needed to directly image this process.
\end{itemize}

\begin{acknowledgments}
{\bf Acknowledgments:} We thank the anonymous referee for reading the paper carefully and providing comments that helped improve and strengthen this paper. 
We would like to also thank Helmut Dannerbauer for his contribution for obtaining the submillimeter data for the series of this work. 
Z. C., S. Z., Y. W., and M. L. are supported by the National Key R\&D Program of China (grant No. 2018YFA0404503), the National Science Foundation of China (grant No. 12073014),  and Tsinghua University Initiative Scientific Research Program (No. 20223080023).  
X. D. and S. W. are supported by Tsinghua University Initiative Scientific Research Program (grant No. 2019Z07L02017). 
This work is based on observations made with the Keck-II Telescope. 
The data were obtained from Keck Observatory Archive.
\end{acknowledgments}

%

\vspace{5mm}
\facilities{Keck-II (KCWI)}


\software{\textsc{astropy} \citep{astropy2022},  
          \textsc{Cloudy} \citep{Ferland2017}, \textsc{vorbin} \citep{Cappellari2003}, \textsc{Linetools} \citep{linetools2017}}

\begin{table*}[b!t]
\centering
\begin{tabular}{l|cccccccc} 
\hline
\hline
\multicolumn{9}{c}{Ly$\alpha$}                                                                                                                                                                                                              \\ 
\hline
\multirow{2}{*}{}                  & \multicolumn{5}{c}{Absoprtion}                                                                                & \multicolumn{3}{c}{Emission}                                                           \\ 
\cline{2-9}
                                   & $\lambda_{\rm abs}$ [${\rm \AA}$] & $N_{\rm HI}$ [cm$^{-2}$] & $b$ [km s$^{-1}$] & $f$    & $\Gamma$ [$\times 10^{8}$ s$^{-1}$] & $\lambda_{\rm emis}$ [${\rm \AA}$] & $A$ [erg s$^{-1}$ cm$^{-2}$ ${\rm \AA^{-1}}$] & $\sigma$ [km s$^{-1}$]  \\ 
\hline
\multicolumn{1}{c|}{initial value} & 4434                      & $10^{16}$              & 100             & 0.4162 & 6.258                           & 4434                       & $F_{\rm peak}$                      & 300                   \\
\multicolumn{1}{c|}{range}         & [4419, 4449]            & [$10^{12}$, $10^{20}$] & [45, 200]     & 0.4162 & 6.258                           & [4419, 4449]             & [100\%, 200\%]$F_{\rm peak}$      & [100, 600]          \\
\hline
\hline
\end{tabular}
\caption{The initial value and range of fitting parameters for Ly$\alpha$. The initial value of the wavelength corresponds to the systemic redshift.The wavelength range corresponds to -1000 km s$^{-1}\sim$ 1000 km s$^{-1}$ centering on the line center. $F_{\rm peak}$ denotes the peak flux in the spectrum extracted from each spatial bin.}
\label{lya_fittings}
\end{table*}

\newpage

\begin{table*}[b!t]
\begin{adjustwidth}{-1.0cm}{}
\centering
\begin{tabular}{l|cccccccc} 
\hline\hline
\multicolumn{9}{c}{CIV}                                                                                                                                                                                                                              \\ 
\hline
\multirow{2}{*}{}                  & \multicolumn{5}{c}{Absoprtion}                                                                                         & \multicolumn{3}{c}{Emission}                                                           \\ 
\cline{2-9}
                                   & $\lambda_{\rm abs}$ [$\rm \AA$] & $N_{\rm CIV}$ [cm$^{-2}$] & $b$ [km s$^{-1}$] & $f$            & $\Gamma$ [$\times 10^{8}$ s$^{-1}$] & $\lambda_{\rm emis}$ [$\rm \AA$] & $A$ [erg s$^{-1}$ cm$^{-2}$ $\AA^{-1}$] & $\sigma$ [km s$^{-1}$]  \\ 
\hline
\multicolumn{1}{c|}{initial value} & 5646, 5653                & $10^{14.5}$             & 100             & 0.1899, 0.0948 & 2.643, 2.628                    & 5646, 5653                 & $F_{\rm peak}$                      & 300                   \\
\multicolumn{1}{c|}{range}         & [5627, 5671]            & [$10^{13}$, $10^{16}$]  & [45, 200]     & 0.1899, 0.0948 & 2.643, 2.628                    & [5627, 5671]             & [100\%, 200\%]$F_{\rm peak}$      & [100, 600]          \\
\hline\hline
\end{tabular}
\caption{The initial value and range of fitting parameters for \civ. 
Since \civ \ is a doublet, two initial values are given corresponding to the doublet of $1548\AA$ and $1550\AA$. 
The lower limit of the wavelength range corresponds to -1000 km s$^{-1}$ centering on $\lambda=5646\AA$ and the upper limit of the wavelength range corresponds to 1000 km s$^{-1}$ centering on $\lambda=5653\AA$.}
\label{civ_fittings}
\end{adjustwidth}
\end{table*}

\begin{table*}[b!t]
\centering
\begin{tabular}{l|lcc} 
\hline
\hline
\multicolumn{4}{c}{HeII}                                                        \\ 
\hline
\multirow{2}{*}{}                  & \multicolumn{3}{c}{Emission}               \\ 
\cline{2-4}
                                   & $\lambda_{\rm emis}$ [$\AA$] & $A$ [erg s$^{-1}$ cm$^{-2}$ $\AA^{-1}$] & $\sigma$ [km s$^{-1}$]  \\ 
\hline
\multicolumn{1}{c|}{initial value} &       5983    & $F_{\rm peak}$  & 300       \\
\multicolumn{1}{c|}{range}         &     [5963, 6003]  & [100\%, 200\%]$F_{\rm peak}$   &   [100, 600]     \\
\hline
\hline
\end{tabular}
\caption{The initial value and range of fitting parameters of \heii. 
The setting is the same as the Ly$\alpha$.}
\label{heii_fitting}
\end{table*}

\begin{table*}[b!t]
\begin{adjustwidth}{-1.3cm}{}
\begin{tabular}{ccc|ccc|c}
\hline
\hline
\multicolumn{3}{c|}{Ly$\alpha$}                    & \multicolumn{3}{c|}{CIV}                    & HeII             \\ \hline
log(N$_{\rm HI}$/cm$^{-2}$)  & $\lambda_{\rm abs, Ly\alpha}$ [$\AA$] & $\lambda_{\rm emis, Ly\alpha}$ [$\AA$] & log(N$_{\rm CIV}$/cm$^{-2}$)  & $\lambda_{\rm abs, CIV}$ [$\AA$] & $\lambda_{\rm emis, CIV}$ [$\AA$] & $\lambda_{\rm emis, HeII}$ [$\AA$] \\ \hline
14.4$\pm$0.2 & 4433.1$\pm$0.2     & 4435.0$\pm$0.1      & 14.9$\pm$0.2 & 5647.7$\pm$0.5     & 5649.1$\pm$0.1      & 5981.9$\pm$0.4       \\
14.4$\pm$0.2 & 4433.2$\pm$0.2     & 4435.1$\pm$0.1      & 14.8$\pm$0.4 & 5647.1$\pm$0.6     & 5649.1$\pm$0.1      & 5982.1$\pm$0.3       \\
14.5$\pm$0.2 & 4433.3$\pm$0.3     & 4436.2$\pm$0.1      & 14.9$\pm$0.2 & 5647.1$\pm$0.4     & 5648.8$\pm$0.2      & 5981.4$\pm$0.3       \\
14.4$\pm$0.2 & 4433.4$\pm$0.2     & 4436.2$\pm$0.1      & 14.9$\pm$0.2 & 5647.5$\pm$0.5     & 5648.4$\pm$0.2      & 5982.0$\pm$0.3       \\
14.2$\pm$0.2 & 4433.0$\pm$0.2     & 4435.4$\pm$0.3      & 14.8$\pm$0.4 & 5647.4$\pm$0.5     & 5648.6$\pm$0.1      & 5982.8$\pm$0.3       \\
14.2$\pm$0.2 & 4432.6$\pm$0.4     & 4436.2$\pm$0.0      & 14.8$\pm$0.2 & 5647.4$\pm$0.5     & 5648.4$\pm$0.1      & 5982.1$\pm$0.3       \\
14.0$\pm$0.2 & 4433.1$\pm$0.3     & 4436.1$\pm$0.1      & 14.9$\pm$0.0 & 5647.7$\pm$0.3     & 5648.2$\pm$0.1      & 5981.8$\pm$0.4       \\
14.2$\pm$0.2 & 4432.9$\pm$0.3     & 4435.4$\pm$0.2      & 14.8$\pm$0.4 & 5647.3$\pm$0.5     & 5648.7$\pm$0.1      & 5983.2$\pm$0.3       \\
14.2$\pm$0.2 & 4433.1$\pm$0.3     & 4435.9$\pm$0.1      & 14.8$\pm$0.4 & 5648.2$\pm$0.3     & 5648.0$\pm$0.1      &          \\
14.6$\pm$0.2 & 4432.9$\pm$0.3     & 4434.6$\pm$0.1      & 14.7$\pm$0.4 & 5648.1$\pm$0.4     & 5649.9$\pm$0.1      &          \\
14.2$\pm$0.2 & 4433.2$\pm$0.3     & 4435.9$\pm$0.1      &  &        &         &          \\
14.7$\pm$0.3 & 4433.1$\pm$0.4     & 4435.9$\pm$0.1      &  &        & 5650.6$\pm$0.6      &          \\
14.3$\pm$0.0 & 4432.3$\pm$0.2     & 4435.2$\pm$0.1      & 14.8$\pm$0.4 & 5647.5$\pm$0.4     & 5648.4$\pm$0.1      & 5982.9$\pm$0.3       \\
14.3$\pm$0.3 & 4432.3$\pm$0.3     & 4435.1$\pm$0.2      & 13.9$\pm$0.4 & 5647.1$\pm$1.2     & 5651.6$\pm$0.2      &          \\
14.5$\pm$0.2 & 4431.4$\pm$0.5     & 4435.0$\pm$0.1      & 14.8$\pm$0.2 & 5647.8$\pm$0.3     & 5648.5$\pm$0.2      &          \\
14.4$\pm$0.2 & 4432.4$\pm$0.3     & 4434.7$\pm$0.1      & 14.8$\pm$0.3 & 5647.2$\pm$0.5     & 5648.4$\pm$0.1      & 5983.1$\pm$0.3       \\
13.6$\pm$0.2 & 4433.5$\pm$0.2     & 4435.6$\pm$0.1      &  &        &         &          \\
13.7$\pm$0.2 & 4433.0$\pm$0.6     & 4435.6$\pm$0.1      & 14.7$\pm$0.2 & 5647.9$\pm$0.5     & 5648.1$\pm$0.4      &          \\
14.4$\pm$0.3 & 4431.9$\pm$0.5     & 4434.7$\pm$0.2      & 14.9$\pm$0.1 & 5647.1$\pm$0.4     & 5648.4$\pm$0.1      & 5983.6$\pm$0.2       \\
14.2$\pm$0.2 & 4432.3$\pm$0.5     & 4436.2$\pm$0.1      & 14.3$\pm$0.1 & 5646.5$\pm$0.3     & 5651.5$\pm$0.2      & 5983.7$\pm$0.3       \\
 &        & 4433.4$\pm$0.4      &  &        &         &          \\
14.2$\pm$0.2 & 4433.5$\pm$0.2     & 4435.8$\pm$0.1      & 14.8$\pm$0.4 & 5649.2$\pm$0.8     & 5645.0$\pm$0.0      & 5978.2$\pm$1.1       \\
 &        & 4435.0$\pm$0.2      &  &        &         &          \\
14.1$\pm$0.2 & 4430.1$\pm$0.6     & 4432.8$\pm$0.6      &  &        &         &          \\
14.4$\pm$0.1 & 4429.8$\pm$0.6     & 4433.9$\pm$0.1      &  &        & 5648.4$\pm$0.5      &          \\
14.2$\pm$0.3 & 4430.2$\pm$0.7     & 4434.1$\pm$0.1      & 14.5$\pm$0.2 & 5647.2$\pm$0.4     & 5647.6$\pm$0.2      &          \\
14.4$\pm$0.2 & 4429.4$\pm$0.7     & 4433.8$\pm$0.1      & 14.6$\pm$0.4 & 5646.4$\pm$0.8     & 5647.2$\pm$0.2      &          \\
14.3$\pm$0.0 & 4431.0$\pm$0.3     & 4435.7$\pm$0.1      & 13.8$\pm$0.3 & 5646.8$\pm$0.8     & 5651.4$\pm$0.2      & 5983.8$\pm$0.3       \\
14.2$\pm$0.2 & 4433.4$\pm$0.1     & 4436.1$\pm$0.1      &  &        & 5648.8$\pm$0.4      &          \\
14.1$\pm$0.1 & 4433.4$\pm$0.2     & 4436.0$\pm$0.1      &  &        & 5646.2$\pm$0.3      &          \\
14.7$\pm$0.3 & 4433.2$\pm$0.3     & 4436.4$\pm$0.2      & 14.6$\pm$0.2 & 5648.0$\pm$0.4     & 5649.4$\pm$0.2      &          \\
 &        & 4431.1$\pm$0.5      &  &        &         &          \\
14.8$\pm$0.2 & 4433.4$\pm$0.2     & 4436.0$\pm$0.1      & 14.9$\pm$0.0 & 5647.8$\pm$0.2     & 5648.0$\pm$0.1      & 5981.8$\pm$0.4       \\ \hline \hline
\end{tabular}
\end{adjustwidth}
\caption{The fitting parameters of the Ly$\alpha$, \heii, and \civ \ lines. 
The log($\rm N_{\rm HI}$/cm$^{-2}$), log($\rm N_{\rm CIV}$/cm$^{-2}$), $\lambda_{\rm abs, Ly\alpha}$, and $\lambda_{\rm abs, CIV}$ are from fitting the absorption. 
The $\lambda_{\rm emis, Ly\alpha}$, $\lambda_{\rm emis, CIV}$, and $\lambda_{\rm emis, HeII}$ are from fitting the emissions. 
The blanks in the column are the parameter that is neglected according to the selection criteria from the Sec.~\ref{spectra_analysis}.}
\label{fitting_table}
\end{table*}

\begin{table*}[b!t]
\begin{adjustwidth}{-1.3cm}{}
\begin{tabular}{ccc|ccc|c}
\hline
\hline
\multicolumn{3}{c|}{Ly$\alpha$}                    & \multicolumn{3}{c|}{CIV}                    & HeII             \\ \hline
log(N$_{\rm HI}$/cm$^{-2}$)  & $\lambda_{\rm abs, Ly\alpha}$ [$\AA$] & $\lambda_{\rm emis, Ly\alpha}$ [$\AA$] & log(N$_{\rm CIV}$/cm$^{-2}$)  & $\lambda_{\rm abs, CIV}$ [$\AA$] & $\lambda_{\rm emis, CIV}$ [$\AA$] & $\lambda_{\rm emis, HeII}$ [$\AA$] \\ \hline
14.8$\pm$0.2 & 4433.4$\pm$0.2     & 4436.0$\pm$0.1      & 14.9$\pm$0.0 & 5647.8$\pm$0.2     & 5648.0$\pm$0.1      & 5981.8$\pm$0.4       \\
14.5$\pm$0.2 & 4433.4$\pm$0.2     & 4436.1$\pm$0.2      & 14.2$\pm$0.3 & 5647.4$\pm$0.8     & 5652.8$\pm$0.3      & 5981.3$\pm$0.4       \\
14.8$\pm$0.2 & 4433.1$\pm$0.2     & 4436.4$\pm$0.1      & 14.9$\pm$0.0 & 5647.1$\pm$0.3     & 5649.2$\pm$0.2      & 5981.6$\pm$0.4       \\
14.2$\pm$0.2 & 4433.3$\pm$0.2     & 4436.1$\pm$0.2      & 14.8$\pm$0.4 & 5647.6$\pm$0.3     & 5649.2$\pm$0.1      & 5983.0$\pm$0.4       \\
 &        & 4432.5$\pm$0.3      &  &        &         &          \\
14.4$\pm$0.2 & 4429.4$\pm$0.8     & 4434.0$\pm$0.1      & 14.6$\pm$0.0 & 5645.4$\pm$0.5     & 5647.2$\pm$0.1      & 5983.8$\pm$0.4       \\
 &        & 4433.0$\pm$0.1      &  &        &         &          \\
 &        & 4433.6$\pm$0.2      &  &        & 5646.5$\pm$0.8      &          \\
14.0$\pm$0.2 & 4433.5$\pm$0.2     & 4436.4$\pm$0.3      &  &        & 5646.2$\pm$0.4      & 5984.5$\pm$0.9       \\
14.1$\pm$0.2 & 4433.5$\pm$0.1     & 4436.2$\pm$0.3      & 14.5$\pm$0.1 & 5650.0$\pm$0.5     & 5648.7$\pm$0.4      &          \\
14.1$\pm$0.2 & 4433.5$\pm$0.2     & 4436.0$\pm$0.2      & 14.7$\pm$0.3 & 5648.2$\pm$0.4     & 5648.3$\pm$0.2      & 5985.0$\pm$0.5       \\
 &        & 4437.9$\pm$0.4      &  &        &         &          \\
14.0$\pm$0.2 & 4432.5$\pm$0.6     & 4432.7$\pm$0.1      &  &        &         &          \\
14.4$\pm$0.2 & 4429.5$\pm$1.0     & 4434.0$\pm$0.1      & 14.5$\pm$0.0 & 5645.2$\pm$0.4     & 5647.3$\pm$0.1      & 5984.0$\pm$0.5       \\
14.3$\pm$0.3 & 4433.5$\pm$0.2     & 4436.8$\pm$0.1      & 14.8$\pm$0.2 & 5647.2$\pm$1.3     & 5648.8$\pm$0.2      & 5983.7$\pm$0.5       \\
14.5$\pm$0.2 & 4433.3$\pm$0.2     & 4436.3$\pm$0.1      & 14.9$\pm$0.2 & 5647.3$\pm$0.4     & 5649.4$\pm$0.1      & 5981.6$\pm$0.4       \\
14.4$\pm$0.3 & 4430.7$\pm$0.8     & 4434.9$\pm$0.1      & 13.7$\pm$0.3 & 5647.1$\pm$0.9     & 5651.2$\pm$0.2      & 5984.1$\pm$0.3       \\
14.3$\pm$0.3 & 4432.1$\pm$0.4     & 4435.8$\pm$0.1      & 14.8$\pm$0.2 & 5647.7$\pm$0.3     & 5649.1$\pm$0.1      & 5983.9$\pm$0.3       \\
14.4$\pm$0.2 & 4431.2$\pm$0.6     & 4434.1$\pm$0.0      & 14.1$\pm$0.3 & 5646.7$\pm$1.6     & 5649.1$\pm$0.2      & 5983.4$\pm$0.6       \\
14.3$\pm$0.2 & 4433.6$\pm$0.3     & 4436.6$\pm$0.1      & 14.5$\pm$0.1 & 5647.3$\pm$0.4     & 5651.9$\pm$0.3      & 5983.5$\pm$0.4       \\
 &        & 4438.9$\pm$0.2      &  &        &         &          \\
14.2$\pm$0.2 & 4433.2$\pm$0.2     & 4435.9$\pm$0.2      & 14.5$\pm$0.3 & 5648.4$\pm$0.4     & 5649.9$\pm$0.1      & 5983.6$\pm$0.4       \\
14.2$\pm$0.3 & 4433.4$\pm$0.1     & 4436.1$\pm$0.2      & 14.1$\pm$0.3 & 5646.8$\pm$1.4     & 5652.6$\pm$0.2      & 5982.5$\pm$0.4       \\
14.2$\pm$0.3 & 4433.4$\pm$0.2     & 4435.0$\pm$0.4      & 14.8$\pm$0.4 & 5647.5$\pm$0.4     & 5649.6$\pm$0.2      & 5983.2$\pm$0.5       \\
14.4$\pm$0.2 & 4431.6$\pm$0.5     & 4435.1$\pm$0.1      & 14.5$\pm$0.2 & 5649.4$\pm$0.5     & 5650.0$\pm$0.1      &          \\
14.3$\pm$0.1 & 4431.6$\pm$0.4     & 4435.3$\pm$0.1      & 14.5$\pm$0.2 & 5648.5$\pm$0.4     & 5649.3$\pm$0.2      & 5983.9$\pm$1.5       \\
14.4$\pm$0.3 & 4433.2$\pm$0.2     & 4434.7$\pm$0.1      & 14.6$\pm$0.2 & 5649.3$\pm$0.3     & 5649.8$\pm$0.1      & 5982.3$\pm$0.6       \\
14.3$\pm$0.2 & 4433.0$\pm$0.2     & 4435.6$\pm$0.2      & 14.0$\pm$0.2 & 5648.5$\pm$1.2     & 5652.6$\pm$0.2      & 5984.6$\pm$0.5       \\
14.2$\pm$0.2 & 4433.5$\pm$0.7     & 4437.4$\pm$0.1      & 14.8$\pm$0.0 & 5647.5$\pm$0.3     & 5651.0$\pm$0.2      & 5984.6$\pm$0.6       \\
 &        &         &  &        &         &          \\
 &        & 4432.6$\pm$0.3      &  &        &         &          \\
14.3$\pm$0.2 & 4433.0$\pm$0.2     & 4435.8$\pm$0.7      & 14.1$\pm$0.3 & 5646.8$\pm$2.0     & 5653.2$\pm$0.4      &          \\
14.2$\pm$0.2 & 4430.8$\pm$0.7     & 4434.6$\pm$0.3      &  &        &         &          \\
 &        & 4432.7$\pm$0.1      &  &        &         &          \\
 &        & 4437.6$\pm$0.2      &  &        &         &          \\
14.3$\pm$0.3 & 4431.7$\pm$0.4     & 4434.6$\pm$0.3      &  &        &         &          \\
 &        & 4436.1$\pm$0.5      &  &        &         &          \\ \hline \hline
\end{tabular}
\end{adjustwidth}
\caption{Continued}
\label{fitting_table2}
\end{table*}

\begin{table*}
\centering
\begin{tabular}{c|ccc}
\hline
\hline
 & Ly$\alpha$ & \civ & \heii  \\ 
 \hline
 k [km s$^{-1}$ kpc$^{-1}$]  &        $8.3\pm3.4$        &    $4.2\pm2.3$        &  $1.6\pm3.3$   \\
 \hline
 \hline
\end{tabular}
\caption{The best-fitting slopes of the velocity profile with the radius of $10 \ {\rm kpc}\leq R \leq 30 \ {\rm kpc}$. 
The slopes of the Ly$\alpha$ and \civ \ have positive values while the slope of \heii \ is around zero.}
\label{vslope}
\end{table*}

\begin{table*}[b!t]
\centering
\label{model_fitting}
\begin{tabular}{c|cc} 
\hline
\hline
           & initial value & range  \\ 
\hline
$r_{\rm init}$ [kpc]   &        $r_{\rm vir}$        &        [0, 3$r_{\rm vir}$]         \\
$v_{\rm r,init}$ [km s$^{-1}$]  &        0.00        &       [-800, +800]          \\
$v_{\rm t,init}$  [km s$^{-1}$]&         0.00       &        [-800, +800]         \\
${\rm log}(m_{\rm cool}/M_{\odot})$ &    5.00            &      [2, 8]           \\
$\alpha$ [Gyr$^{-1}$]     &        1.71        &        [0, 10]         \\
$\theta$ [rad]    &       $\pi/2$         &         [0, $\pi$]         \\
$f_{\rm hot}$     &       0.35         &          [0, 1]      \\
\hline
\hline
\end{tabular}
\caption{The initial value and prior of free parameters input into the MCMC. 
Quiet large priors are given to the seven free parameters. 
For the initial radial velocity and tangential velocity, the minus means the negative direction.}
\label{parameter_boundary}
\end{table*}










\bibliography{sample631}{}
\bibliographystyle{aasjournal}



\end{document}